\documentclass[prd,amsmath,aps,floats,amssymb, floatfix, superscriptaddress,nofootinbib,preprintnumbers,twocolumn]{revtex4-1}  
\usepackage{lineno} 

\usepackage{amsmath,amssymb,natbib,latexsym,times}
\usepackage{graphics}
\usepackage{todonotes,multirow,enumitem}

\newcommand{\sfig}[2]{
\includegraphics[width=#2]{#1}
        }
\newcommand{\Sfig}[2]{
    \begin{figure}[t]
    \sfig{Figures/#1.pdf}{\columnwidth}
    \caption{{\small #2}}
    \label{fig:#1}
    \end{figure}
}
\newcommand{\Swide}[2]{
\begin{figure*}[thbp]
 \sfig{Figures/#1.pdf}{.8\textwidth}
  \caption{{\small #2}}
   \label{fig:#1}
   \end{figure*}
}

\newcommand{\Neff}{N_{\mathrm{eff}}}
\newcommand\salpha{{\eta_{\rm IA}}}

\newcommand\be{\begin{equation}}
\newcommand\ee{\end{equation}}
\def\hinvmpc{\,h^{-1}{\rm Mpc}}

\newcommand\dchisq{[$(\Delta \chi^2)_{\rm DES},(\Delta \chi^2)_{\rm Ext},(\Delta \chi^2)_{\rm DES+Ext}$]}

\newcommand\cosmosis{{\tt CosmoSIS}}
\newcommand\cosmolike{{\tt CosmoLike}}
\newcommand\metacal{{\textsc{metacalibration}}}

\begin{document}

\title{Dark Energy Survey Year 1 Results:\\Constraints on Extended Cosmological Models from
  Galaxy Clustering and Weak Lensing}


\author{T.~M.~C.~Abbott}
\affiliation{Cerro Tololo Inter-American Observatory, National Optical Astronomy Observatory, Casilla 603, La Serena, Chile}
\author{F.~B.~Abdalla}
\affiliation{Department of Physics \& Astronomy, University College London, Gower Street, London, WC1E 6BT, UK}
\affiliation{Department of Physics and Electronics, Rhodes University, PO Box 94, Grahamstown, 6140, South Africa}
\author{S.~Avila}
\affiliation{Institute of Cosmology \& Gravitation, University of Portsmouth, Portsmouth, PO1 3FX, UK}
\author{M.~Banerji}
\affiliation{Kavli Institute for Cosmology, University of Cambridge, Madingley Road, Cambridge CB3 0HA, UK}
\affiliation{Institute of Astronomy, University of Cambridge, Madingley Road, Cambridge CB3 0HA, UK}
\author{E.~Baxter}
\affiliation{Department of Physics and Astronomy, University of Pennsylvania, Philadelphia, PA 19104, USA}
\author{K.~Bechtol}
\affiliation{LSST, 933 North Cherry Avenue, Tucson, AZ 85721, USA}
\author{M.~R.~Becker}
\affiliation{Argonne National Laboratory, 9700 South Cass Avenue, Lemont, IL 60439, USA}
\author{E.~Bertin}
\affiliation{CNRS, UMR 7095, Institut d'Astrophysique de Paris, F-75014, Paris, France}
\affiliation{Sorbonne Universit\'es, UPMC Univ Paris 06, UMR 7095, Institut d'Astrophysique de Paris, F-75014, Paris, France}
\author{J.~Blazek}
\affiliation{Center for Cosmology and Astro-Particle Physics, The Ohio State University, Columbus, OH 43210, USA}
\affiliation{Institute of Physics, Laboratory of Astrophysics, \'Ecole Polytechnique F\'ed\'erale de Lausanne (EPFL), Observatoire de Sauverny, 1290 Versoix, Switzerland}
\author{S.~L.~Bridle}
\affiliation{Jodrell Bank Center for Astrophysics, School of Physics and Astronomy, University of Manchester, Oxford Road, Manchester, M13 9PL, UK}
% ...

\collaboration{DES Collaboration}

\date{\today}

\label{firstpage}
\begin{abstract}
We present constraints on extensions of the minimal cosmological models
dominated by dark matter and dark energy, $\Lambda$CDM and $w$CDM, by using a
combined analysis of galaxy clustering and weak gravitational lensing from the
first-year data of the Dark Energy Survey (DES Y1) in combination with
external data. We consider four extensions of the minimal dark
energy-dominated scenarios: 1) nonzero curvature $\Omega_k$, 2) number of
relativistic species $\Neff$ different from the standard value of 3.046, 3)
time-varying equation-of-state of dark energy described by the parameters
$w_0$ and $w_a$ (alternatively quoted by the values at the pivot redshift,
$w_p$, and $w_a$), and 4) modified gravity described by the parameters $\mu_0$
and $\Sigma_0$ that modify the metric potentials. We also consider external
information from Planck cosmic microwave background measurements; baryon
acoustic oscillation measurements from SDSS, 6dF, and BOSS; redshift-space
distortion measurements from BOSS; and type Ia supernova information from the
Pantheon compilation of datasets. Constraints on curvature and the number of
relativistic species are dominated by the external data; when these are
combined with DES Y1, we find $\Omega_k=0.0020^{+0.0037}_{-0.0032}$ at the 68\%
confidence level, and the upper limit $\Neff<3.28\, (3.55)$ at 68\%
(95\%) confidence, assuming a hard prior $\Neff>3.0$. For the
time-varying equation-of-state, we find the pivot value $(w_p,
w_a)=(-0.91^{+0.19}_{-0.23}, -0.57^{+0.93}_{-1.11})$ at pivot redshift
$z_p=0.27$ from DES alone, and $(w_p, w_a)=(-1.01^{+0.04}_{-0.04},
-0.28^{+0.37}_{-0.48})$ at $z_p=0.20$ from DES Y1 combined with external data; in
either case we find no evidence for the temporal variation of the equation of
state. For modified gravity, we find the present-day value of the relevant
parameters to be $\Sigma_0= 0.43^{+0.28}_{-0.29}$ from DES Y1 alone, and
$(\Sigma_0, \mu_0)=(0.06^{+0.08}_{-0.07}, -0.11^{+0.42}_{-0.46})$ from DES Y1
combined with external data. These modified-gravity constraints are
consistent with predictions from general relativity.
\end{abstract}

\preprint{DES-2018-0376}
\preprint{FERMILAB-PUB-18-507-PPD}
\maketitle

\section{Introduction}
\label{sec:intro}

Evidence for dark matter \cite{Zwicky:1933gu} and the discovery of cosmic
acceleration and thus evidence for dark energy
\cite{Riess:1998cb,Perlmutter:1998np} were pinnacle achievements of cosmology
in the 20th century. Yet because of the still-unknown physical mechanisms
behind these two components, understanding them presents a grand challenge for
the present-day generation of cosmologists. Dark matter presumably corresponds
to an as-yet undiscovered elementary particle whose existence, along with
couplings and other quantum properties, is yet to be confirmed and
investigated. Dark energy is even more mysterious, as there are no compelling
models aside, arguably, from the simplest one of vacuum energy.
 
Dark matter and dark energy leave numerous unambiguous
imprints in the expansion rate of the universe and in the rate of growth of
cosmic structures as a function of time. The theoretical modeling and direct
measurements of these signatures have led to a renaissance in data-driven
cosmology.  Numerous ground- and space-based sky surveys have dramatically
improved our census of dark matter and dark energy over the past two decades,
and have led to a consensus model with $\sim$5\% energy density in baryons,
$\sim$25\% in cold (nonrelativistic) dark matter (CDM), and $\sim$70\% in
dark energy. These probes, reviewed in \cite{Frieman:2008sn,Weinberg:2012es,Huterer:2017buf},
include the cosmic microwave background (CMB; \cite{Hu:2001bc}); galaxy clustering including the location of the baryon acoustic oscillation (BAO) feature and the impact of redshift space distortions (RSD); distances to type Ia supernovae (SNe Ia); 
weak gravitational lensing (WL \cite{Hoekstra:2008db}), given by tiny distortions in the shapes of galaxies
due to the deflection of light by intervening large-scale structure; and the abundance of clusters of galaxies \cite{Allen:2011zs}.

The simplest and best-known model for dark energy is the
cosmological constant. This model, represented by a single parameter given by
the magnitude of the cosmological constant, is currently in good agreement
with data.  
On the one hand, vacuum energy density is predicted to exist in quantum field theory due to zero-point
energy of quantum oscillators, and manifests itself as a cosmological constant:
unchanging in time and spatially smooth.  On the other hand, the theoretically expected vacuum energy density is
tens of orders of magnitude larger than the observed value as has been known
even prior to the discovery of the accelerating universe
\cite{Weinberg:1988cp,Martin:2012bt}.  Apart from the cosmological constant,
there exists a rich set of other dark energy models including evolving scalar
fields, modifications to general relativity, and other physically-motivated
possibilities \cite{Joyce:2014kja,Bull:2015stt,Ishak:2018his} with many possible avenues to test them
with data \cite{Jain:2007yk}. Testing for such extensions of the simplest dark
energy model on the present-day data has spawned an active research area in
cosmology
\cite{Zhang:2005vt,Caldwell:2007cw,Guzik:2009cm,Bean:2010zq,Zhao:2010dz,Reyes:2010tr,Daniel:2010ky,Zhao:2012aw,Daniel:2012kn,Ade:2015rim,Hojjati:2015ojt,Salvatelli:2016mgy,Joudaki:2016kym,Mueller:2016kpu,Zhao:2017cud,Amon:2017lia,Aghanim:2018eyx},
and is the subject of the present paper.

The Dark Energy Survey (DES\footnote{\url{http://www.darkenergysurvey.org/}})
  \cite{Abbott:2005bi} is a photometric survey imaging the sky in five
filters ($grizY$) using the 570 Mpixels, 3 deg$^2$ field-of-view Dark Energy
Camera (DECam) \cite{DECam}, mounted on the 4-meter Blanco telescope at the Cerro Tololo
International Observatory in Chile. After more than five years of data-taking,
the survey will end in early 2019 with more than 300 million galaxies
catalogued in an area of roughly 5000 deg$^2$.

In 2017 the DES collaboration published the analyses of its first year of data
(Y1). It presented results which put
constraints on certain cosmological parameters derived from their
  late-universe imprints in galaxy surveys at the same level of
  precision as the constraints obtained on these same parameters from
  their \textit{early}-universe signatures in the CMB data. These results, described in \cite{Y1KP} (hereafter Y1KP) are
based on the two-point statistics of galaxy clustering and weak gravitational
lensing. The combined analysis of the three different two-point correlation
functions (galaxy clustering, cosmic shear, and the galaxy-shear
cross-correlation, typically referred to as galaxy-galaxy lensing) is the
end product of a complex set of procedures which includes the analysis
pipeline and methodology \cite{method}, its validation on realistic
simulations \cite{simspaper}, the creation of shape catalogs \cite{shearcat},
the estimation and validation of the redshift distribution for different
galaxy samples \cite{photoz}, measurement and derivation of cosmological
constraints from the cosmic shear signal \cite{shearcorr}, galaxy--galaxy
lensing results \cite{gglpaper} and the galaxy clustering statistics
\cite{wthetapaper}.  Both alone and in combination with external data from CMB
(Planck~\citep{Ade:2015xua}), BAO (6dF Galaxy Survey~\citep{Beutler:2011hx},
the SDSS Data Release 7 Main Galaxy Sample~\citep{Ross:2014qpa}, BOSS Data
Release 12~\citep{Alam:2016hwk}) and SNe Ia (Joint Lightcurve Analysis (JLA
\citep{Betoule:2014frx}), DES provides precise measurements in the parameters
describing the amplitude of mass fluctuations perturbation and the matter
energy density evaluated today. We refer the reader to Y1KP for more details
of the DES Y1 analysis, and to Sec~\ref{sec:extdata} below for further description
  of external data.

In Y1KP we considered only the two simplest models for dark energy: the
standard cosmological constant $\Lambda$CDM model and a $w$CDM model with an
extra parameter (the dark energy equation-of-state $w$) accounting for a
constant relation between the pressure and the energy density of the dark
energy fluid ($p= w\rho$).  In this paper we explore the impact of the
DES Y1 data on the analysis of a few extensions of the standard flat
$\Lambda$CDM and $w$CDM models considered in Y1KP, namely  the possibilities of:
\begin{itemize}
\item Nonzero spatial curvature; 
\item New relativistic degrees of freedom; 
\item Time-variation of the dark energy equation-of-state;
\item Modifications of the laws of gravity on cosmological scales.
\end{itemize}
We describe these extensions in more detail below.

Our analysis applies the same validation tests with respect to assumptions
about the systematic biases, analysis choices, and pipeline accuracy, as
previously done in Y1KP. We also adopt the parameter-level blinding procedure
used in that paper, and we do not look at the final cosmological constraints
until after unblinding, when the analysis procedure and estimates of
uncertainties on various measurement and astrophysical nuisance parameters
were frozen. Validation and parameter blinding are also described in further
detail below.

Our study effectively complements and extends a number of studies of
extensions to $\Lambda$/$w$CDM in the literature using state-of-the-art data,
e.g. by Planck \cite{Ade:2015xua,Ade:2015rim}, the Baryon Oscillation
Spectroscopic Survey (BOSS) \cite{Alam:2016hwk}, the Kilo Degree Survey (KiDS)
\cite{Joudaki:2016kym,Joudaki:2017zdt} and more recently by using the Pantheon
compilation of SNe Ia data \cite{Scolnic:2017caz}. These studies report no
significant deviations from $\Lambda$CDM. We will comment on the comparison of
our results to these existing constraints in the conclusions.

The paper is organized as follows: the data sets used in the analyses are
described in \S\ref{sec:data}, while the models and parameters used to
describe the data are detailed in \S\ref{sec:theory}. To ensure that our
analysis will not misattribute an astrophysical systematic error to a {\it
  detection} of an extension, we present a series of validation tests in
\S\ref{sec:valid}. In \S\ref{sec:results}, we present our results before
concluding in \S\ref{sec:conclu}.

\section{Data}
\label{sec:data}

The primary data used in this study are the auto- and cross-correlations of galaxy positions and shapes measured in data taken by the Dark Energy Survey during its first year of observations.\footnote{The DES Y1 data products used in this work are publicly available from: https://des.ncsa.illinois.edu/releases/y1a1.} We refer the reader to Y1KP for details and only give a summary here. 

\subsection{Catalogs}

The images taken between August 31, 2013 and February 9, 2014 were processed with the DES Data Management (DESDM) system \citep{Desai:2012,Sevilla:2011,Mohr:2008,desdm}, and its outputs validated and filtered to produce the high-quality DES Y1 Gold catalog \citep{Drlica-Wagner:2017tkk}. 

From the galaxies in this catalog, we define two samples to be used here: \emph{lens} galaxies, for which we measure the angular correlation function of positions, and \emph{source} galaxies, for which we measure the autocorrelation of shapes and the cross-correlation of shapes with lens galaxy positions. To reduce the impact of varying survey characteristics and to remove foreground objects and contaminated regions, we define both samples over an area of 1321 deg$^2$.

As \emph{lens} galaxies, we use a sample of luminous red galaxies identified with the \textsc{redMaGiC} algorithm \cite{redmagicSV}. This choice is motivated by the small uncertainties in photometric redshifts, high completeness over most of our survey, and the strong clustering of these galaxies. We divide the \textsc{redMaGiC} sample into five redshift bins, using three different cuts on intrinsic luminosity to ensure completeness. For bins of redshift $z\in[(0.15-0.3), (0.3-0.45), (0.45-0.6)]$, we chose a luminosity cut of $L>0.5L_*$ with a spatial density $\bar n = 10^{-3} (h^{-1} {\rm Mpc})^{-3}$, where the comoving density assumes a fiducial $\Lambda$CDM cosmology. For the additional redshift bins $z\in(0.6-0.75)$ and $(0.75-0.9)$, the luminosity cuts and densities are $L>L_*$, $\bar n = 4\times 10^{-4} (h^{-1} {\rm Mpc})^{-3}$ and $L>1.5L_*$, $\bar n = 10^{-4} (h^{-1} {\rm Mpc})^{-3}$, respectively. In total, these samples contain approximately 660,000 lens galaxies.
 
The primary systematic uncertainties in this catalog are based on residual correlations of galaxy density with observational characteristics of the survey, and in the uncertainty and bias of the lens galaxy redshifts as estimated from the broad-band photometry. The first effect is studied in detail and corrected in \cite{wthetapaper}. The redshift distributions estimated for the \textsc{redMaGiC} galaxies are validated, and the budget for residual uncertainties in quantified, using their clustering with spectroscopic galaxy samples \cite{Cawthon:2017qxi}.

To generate a catalog of \emph{source} galaxies with accurate shapes for estimating lensing signals, we use the \textsc{metacalibration} method \cite{Huff:2017qxu,Sheldon:2017szh} on top of \textsc{ngmix}\footnote{https://github.com/esheldon/ngmix}. \textsc{ngmix} provides the ellipticity measurements for a sufficiently resolved and high signal-to-noise subsample of the Y1 Gold catalog by fitting a simple Gaussian mixture model, convolved with the individual point spread function, to the set of all single exposures taken of a galaxy. The primary systematic uncertainty in this catalog is a multiplicative error on the mean shear measurement due to biases related to noise and selection effects. In the \textsc{metacalibration} scheme, this bias is removed by introducing an artificial shear signal and measuring the \emph{response} of the mean measured ellipticity to the introduced shear. To this end, all galaxy images are artificially sheared, and their ellipticities and all properties used for selecting the sample are remeasured on the sheared versions of their images. By applying a response correction to all estimated shear signals, we find that this method provides measurements with a small multiplicative bias that is dominated by the effect of blending between neighboring galaxies \cite{shearcat}.

To divide these source galaxies into redshift bins, we use the means of the redshift probability distributions provided by a version of the \textsc{BPZ} algorithm \cite{Coe:2006hj}. This procedure is based on the \textsc{metacalibration} measurements of $griz$ galaxy fluxes, as detailed in \cite{photoz}. By splitting on $z_{\rm mean}\in[(0.2-0.43), (0.43-0.63), (0.63-0.9), (0.9-1.3)]$, we generate four bins with approximately equal density. The redshift distribution of each source bin is initially estimated from the stack of individual galaxy BPZ redshift probability distributions. This initial estimate is validated, and the systematic uncertainty on the mean redshift in each bin is estimated using a resampling method of high-quality photometric redshifts gained from multiband data in COSMOS \cite{photoz} and the clustering of the sources with \textsc{redMaGiC} galaxies \cite{Davis:2017dlg, Gatti:2017hmb}.

\Sfig{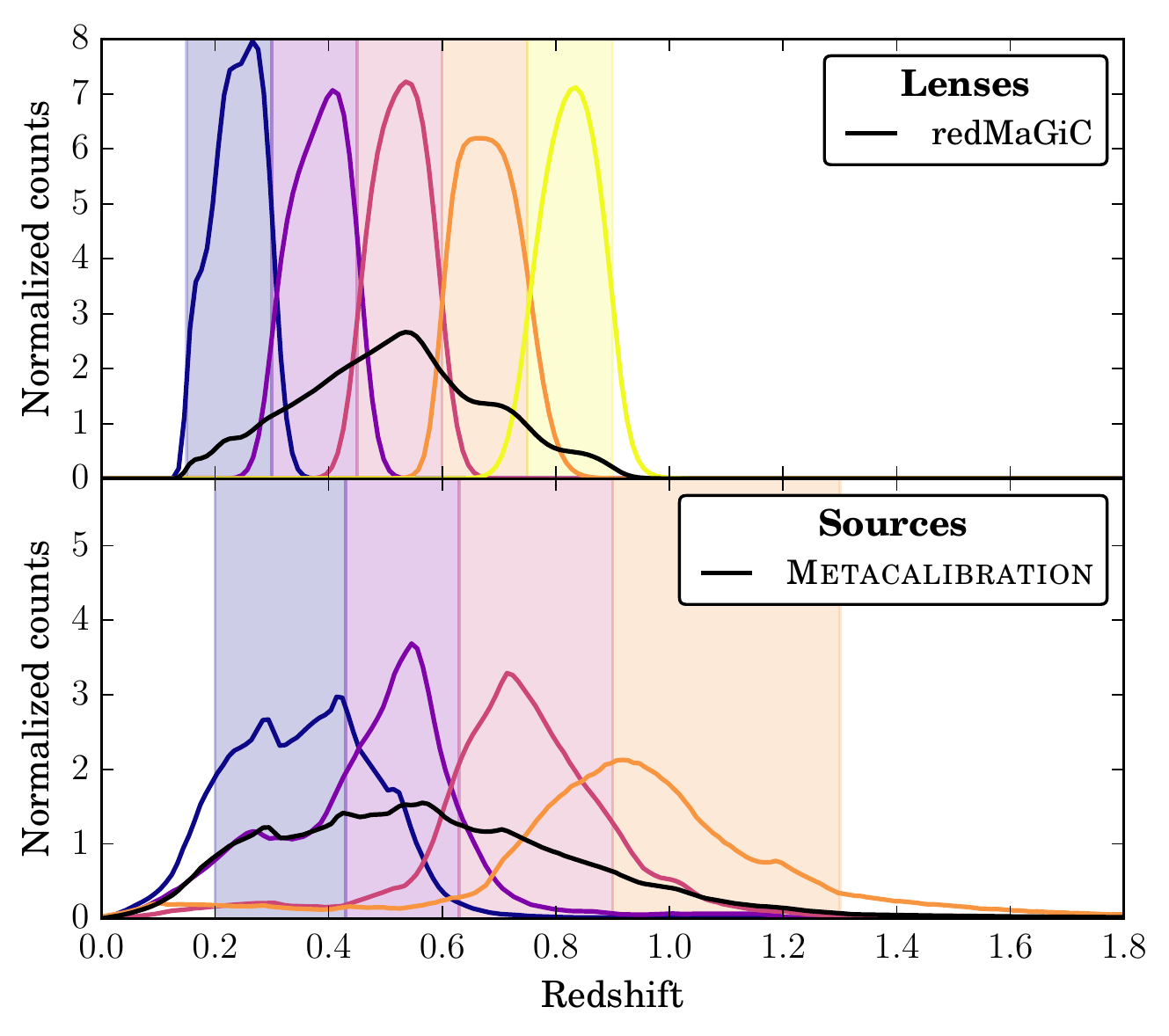}{
  Estimated redshift distributions of the lens and source
  galaxies used in the analysis. The shaded vertical regions define the
  bins: galaxies are placed in the bin spanning their mean
  photo-$z$ estimate. We show both the redshift distributions of galaxies in
  each bin (colored lines) and their overall redshift distributions (black
  lines). }

The lens and source galaxy distributions are shown in Figure~\ref{fig:plot_nofzs_noim3shape}.
The systematic uncertainties on redshift of both samples, and on the shear estimates of the source sample, are quantified in \cite{shearcat,photoz} and marginalized over in all cosmological likelihoods.

\subsection{Measurements}

For the lens and source sample, we use measurements of the three sets of two-point functions in \cite{Y1KP}: 
\begin{itemize}
\item \emph{Galaxy clustering}: the autocorrelation of lens galaxy positions in each redshift bin $w(\theta)$, i.e.~the fractional excess number of galaxy pairs of separation $\theta$ relative to the number of pairs of randomly distributed points within our survey mask \cite{wthetapaper},
\item \emph{Cosmic shear}: the autocorrelation of source galaxy shapes within and between the source redshift bins, of which there are two components $\xi_{\pm}(\theta)$, taking the products of the ellipticity components of pairs of galaxies, either adding ($+$) or subtracting ($-$) the component tangential to the line connecting the galaxies and the component rotated by $\pi/4$ \cite{shearcorr},
\item \emph{Galaxy-galaxy lensing}: the mean tangential ellipticity of source galaxy shapes around lens galaxy positions, for each pair of redshift bins, $\gamma_t(\theta)$ \cite{gglpaper}.
\end{itemize}
 Details of these measurements and the checks for potential systematic effects
 in them are described in detail in \cite{wthetapaper,gglpaper,shearcorr}, and
 an overview of the full data vector is given in \cite{Y1KP}. Here we follow Y1KP,
 and refer to results from combining all 3 two-point functions as ``DES Y1 $3\times2$pt''.

Each of these measurements is performed in a set of 20 logarithmic bins of
angular separation between 2.5' and 250' using the software \textsc{treecorr}
\cite{Jarvis:2003wq}. We only use a subset of these bins, removing small
scales on which our model is not sufficiently accurate. The fiducial
  scales that we use for clustering and galaxy-galaxy lensing correspond to
  minimal scale of $R=8\hinvmpc$ and $12\hinvmpc$, respectively. For cosmic
  shear, the minimal angular scale $\theta_{\rm min}$ is redshift-dependent, and is determined by
  requiring that the cross-correlation $(\xi_{\pm})_{ij}(\theta_{\rm min})$ at a pair of
  redshift bins $i$ and $j$ not incur an expected fractional contribution from
  baryonic interactions exceeding 2\%; see \cite{shearcorr} for details.

For the curvature, number of relativistic species, and dark energy tests, we
use the exact same set of scales as in Y1KP, and the datavector with a total
of 457 measurements in $(w(\theta), \xi_{\pm}(\theta), \gamma_t(\theta))$. For
our modified gravity tests, we use a more stringent range of scales, described
at the end of Sec.~\ref{sec:sigmamu}; this datavector spans only the linear
scales, and has a total of 334 measurements.

DES Y1 measurements provide information at $z\lesssim 1$, when -- in most
models -- dark energy starts to play a role in cosmic evolution. They provide
information about both the geometrical measures (distances, volumes) and the
growth of cosmic structure.  In particular, both lensing and galaxy clustering
are sensitive to the growth of structure, while the kernels in the calculation
of the corresponding two-point correlation functions also encode the geometry
given by distances (see e.g.\ equations in Sec.\ 4 of Y1KP). Therefore, all of
the DES Y1 $3\times2$pt measurements probe both geometry and the growth of
structure, and thus complement the largely geometrical external data discussed
below in Sec.~\ref{sec:extdata}.  The geometry-plus-growth aspect of the DES
Y1 $3\times2$pt measurements makes them particularly sensitive to predictions
of the models studied in this paper such as modified gravity.

\subsection{Covariance}

The statistical uncertainties of these measurements are due to spatial variations in the realizations of the cosmic matter density field (cosmic variance) and random processes governing the positions (shot noise) and intrinsic orientations (shape noise) of galaxies. We describe these uncertainties and their correlations with a covariance matrix $\mathbf{C}$, which is calculated using \cosmolike~\cite{Krause:2016jvl} using the relevant four-point functions in the halo model ~\cite{HaloModel}. Shot and shape noise are scaled according to the actual number of source galaxies in our radial bins to account for source clustering and survey geometry. Details of this approach are described in \cite{Krause:2016jvl,Troxel:2018qll}, along with our validation of the covariance matrix and the corresponding Gaussian likelihood.

\subsection{External data}\label{sec:extdata}

Combining the DES large-scale structure weak lensing and galaxy clustering data
with other, independent probes has benefits in constraining the beyond-minimal
cosmological models considered in this paper. 
In particular, the measurements of distances by SNe Ia and
BAO, along with the distance to recombination from the CMB, provide precise
geometrical measures, while redshift-space distortions (RSD) are
sensitive to the growth of cosmic structure
\cite{Zhang:2003ii,Wang:2007fsa,Abate:2008au,Ruiz:2014hma,Bernal:2015zom}.
These external data significantly complement the combination of geometry and
growth probed by the DES clustering and lensing data. Similarly, combining
DES with external data enables the comparison of the inferred cosmology from early- and late-time
probes (see e.g.\ Fig.\ 11 in Y1KP).

As in Y1KP, we combine DES data with a collection of external data sets to
derive the most precise constraints on the $\Lambda$CDM extensions models. We
use CMB, CMB Lensing, BAO, RSD, and Supernova Ia measurements in various
combinations. Our final set of external data, described in more detail below,
is similar to that used in Y1KP; the main differences are that we add RSD
measurements from BOSS, and that we update the JLA supernova dataset used in
Y1KP to the more recent Pantheon results.

We treat the likelihoods of individual external datasets as independent,
simply summing their log-likelihoods.  We now describe the individual external
datasets that we add to DES data in our combined analysis.

\subsubsection{CMB \& CMB lensing}
The cosmic microwave background
temperature $T$ and polarization ($E$- and $B$-modes) anisotropies are a powerful probe of the
early universe. The combination of a rich phenomenology
with linear perturbations to a background yields very strong constraints on density perturbations
in the early Universe, and on reionization.

In this work we use the Planck 2015 likelihood\footnote{Planck 2018 results \cite{Aghanim:2018eyx}
  were released as this paper was in advanced stages of the analysis, so we
  stick with using the Planck 2015 likelihood. The main difference between the
  two is better measurements of CMB polarization in Planck 2018, resulting in
  better constraints on the optical depth $\tau$. }  as described in \citet{Aghanim:2015xee}.
We use the Planck $TT$ likelihood for multipoles $30 \leq \ell \leq 2508$ and
the joint $TT$, $EE$, $BB$ and $TE$ likelihood for $2 \leq \ell \leq 30$. We
refer to this likelihood combination as TT+lowP.\footnote{We used the public
  Planck likelihood files {\sc plik\_lite\_v18\_TT.clik} and {\sc
    lowl\_SMW\_70\_dx11d\_2014\_10\_03\_v5c\_Ap.clik.}}

Planck primary CMB measurements like these strongly constrain all of the baseline cosmological
parameters that we use across our models.  They have varying power to constrain extension 
parameters.

We also make use of Planck CMB lensing measurements \citep{Ade:2015zua}, from
temperature only\footnote{We use the file {\sc
    smica\_g30\_ftl\_full\_pttptt.clik\_lensing}.}. These are measured from
higher-order correlations in the temperature field, and act like an additional
narrow and very high redshift source sample. We neglect any
  cross-correlation between DES Y1 measurements and Planck's CMB lensing map
  because 1) the noise in the CMB lensing map is sufficiently large; 2) the
  overlap of the surveys is small compared to the total CMB lensing area; and
  3) the CMB lensing autospectrum receives most of its contribution from $z
  \simeq 2$, while DES constraints are at $z \lesssim 1$, which further
  reduces covariance. We have explicitly tested that the
  assumption of ignoring the DES-CMBlens covariance holds to an excellent
  accuracy.

\subsubsection{BAO + RSD}

BAO measurements locate a peak in the correlation function of cosmic structure that corresponds
to the sound horizon at the drag epoch.
Since the sound speed before that point depends only on the well-understood
ratio of photon to baryon density, this horizon acts as a standard ruler and
can be used to measure the angular diameter distance with a percent-level
precision.

As in Y1KP we use BAO measurements from BOSS Data Release 12
\citep{Alam:2016hwk}, which provides measurements of both the Hubble parameter
$H(z_i)$ and the comoving angular diameter distance $d_A(z_i)$, at three
separate redshifts, $z_i = \{0.38, 0.51, 0.61\}$. The other two BAO data that
we use, 6DF Galaxy survey \citep{Beutler:2011hx} and SDSS Data Release 7 Main
Galaxy Sample \citep{Ross:2014qpa}, are lower signal-to-noise and can only
tightly constrain the spherically averaged combination of transverse and
radial BAO modes, $D_V(z) \equiv [c z (1+z)^2 D^2_A(z) / H(z)]^{1/3}$. These
constraints are at respective redshifts $z=0.106$ (6dF) and $z=0.15$ (SDSS
MGS).

We also utilize the redshift-space distortion measurements from BOSS DR12;
they are given as measurements of the quantity $f(z_i)\sigma_8(z_i)$ at the
aforementioned three redshifts. Here $f$ is the linear growth rate of
matter perturbations and $\sigma_8$ is the amplitude of mass fluctuations on
scales $8\hinvmpc$. We employ the full covariance, given by
~\citep{Alam:2016hwk},  between these three RSD measurements
and those of BAO quantities $H(z_i)$ and $d_A(z_i)$. We treat the 6dF and SDSS
MGS measurements as independent of those from BOSS DR12, and we neglect any
cosmological dependence on the derived values of $f(z_i)\sigma_8(z_i)$ from
BOSS DR12 data.

Finally, we ignore the covariance between these BAO/RSD measurements and those of DES
galaxy clustering and weak lensing; the two sets of measurements
are carried out on different areas on the sky and the
covariance is expected to be negligible.

\subsubsection{Supernovae}

Type Ia supernovae (SNe Ia) provide luminosity distances out
to redshift of order unity and beyond, and thus excellent constraints on the
expansion history of the universe. In this analysis we use the Pantheon SNe Ia
sample \cite{Scolnic:2017caz} which combines 279 SNe Ia from the Pan-STARRS1
Medium Deep Survey $(0.03 < z < 0.68)$ with SNe Ia from SDSS, SNLS, various
low-z and HST samples.  The Pantheon data was produced using the Pan-STARRS1
Supercal algorithm \cite{Scolnic:2015eyc} which established global calibration
for 13 different SNe Ia samples.  The final Pantheon sample includes 1048 objects
in the redshift range $0.01 < z < 2.26$.

\section{Theory and modeling}
\label{sec:theory}

\subsection{Standard cosmological parameters}

We assume the same set of $\Lambda$CDM cosmological parameters described in Y1KP, 
then supplement it with parameters alternately
describing four extensions. We parametrize the matter energy density today relative to the critical density
$\Omega_m$, as well as that of the baryons $\Omega_b$ and of neutrinos
$\Omega_{\nu}$\footnote{In the $\Sigma_0$, $\mu_0$ systematic tests that use the older {\tt
  {\tt MGCAMB}}, this was not implemented, so $\Omega_{\nu}$ is fixed in these
  tests. We do vary $\Omega_\nu$ in our runs on the real data.}. Moreover, we adopt the amplitude $A_s$ and
the scalar index $n_s$ of the primordial density perturbations power spectrum,
as well as the optical depth to reionization $\tau$, and the value of the Hubble
parameter today $H_0$. Except in the case of varying curvature, we assume that
the universe is flat and, except in the case of varying dark energy, we assume
that it is $\Lambda$-dominated with $w=-1$; under those two assumptions,
$\Omega_\Lambda=1-\Omega_m$. Note that the amplitude of mass fluctuations
$\sigma_8$ is a derived parameter, as is the parameter that
decorrelates $\sigma_8$ and $\Omega_m$, $S_8\equiv \sigma_8 (\Omega_m/0.3)^{0.5}$.
The fiducial parameter set is therefore
\begin{equation}
  \mathbf{\theta}_{\rm base}=\{\Omega_m, H_0, \Omega_b, n_s, A_s, (\tau)\},
\end{equation}
where the parentheses around the optical depth parameter indicate that it is
used only in the analysis combinations that use CMB data.

To model the fully nonlinear power spectrum, we first estimate the linear
primordial power spectrum on a grid of $(k,z)$ using {\tt
  CAMB}~\cite{Lewis:1999bs} or {\tt CLASS}~\cite{2011arXiv1104.2932L}. We then
apply the {\tt HALOFIT} prescription
\cite{Smith:2002dz,Takahashi:2012em,Bird:2011rb} to get the nonlinear spectrum.
Throughout this work, we employ the version from Takahashi et
al.\ \cite{Takahashi:2012em}.

In addition to this set of $\Lambda$CDM parameters, we use the following
parametrization for each of the extension models:
\begin{enumerate}
\item Spatial curvature: $\Omega_k$; 
\item The effective number of neutrinos species $\Neff$; 
\item Time-varying equation-of-state of dark energy: $w_0$, $w_a$;
\item Tests of gravity: $\Sigma(a)$, $\mu(a)$.
\end{enumerate}
We describe these extensions in more detail below in Sec.~\ref{sec:extensions}.

\subsection{Nuisance parameters}\label{sec:nuisance}

We follow the analysis in Y1KP, and model a variety of systematic
uncertainties using an additional 20 nuisance parameters. 
The nuisance parameters are:
\begin{itemize}
\item Five parameters $b_i$ that model linear bias of lens galaxies in five
  redshift bins;
\item Two parameters, $A_{\rm IA}$ and $\salpha$, that model the power
  spectrum of intrinsic alignments as a power-law scaling $A_{\rm
    IA}(\frac{1+z}{1+z_0})^\salpha$, with $z_0=0.62$ (see Sec.~VIIB of
    \cite{shearcorr} for a complete description of the model);
\item Five parameters $\Delta z_l^i$ to model the uncertainty in the means of
  distributions $n(z_i)$ of galaxies in each of the lens bins;
\item Four parameters  $\Delta z_s^i$ to model the uncertainty in the means of
  distributions $n(z_i)$ of galaxies in each of the source bins;
\item Four parameters $m_i$ that model the overall uncertainty in the multiplicative shear bias in each of the source bins.
\end{itemize}
All of the cosmological and nuisance parameters in our standard analysis,
along with their respective priors, are given in Table \ref{tab:params}.

\begin{table}[t]
\caption{Parameters and priors used to describe the measured two-point
  functions, as adopted from Y1KP. {\it Flat} denotes a flat prior in the range given while {\it
    Gauss}($\mu,\sigma$) is a Gaussian prior with mean $\mu$ and width
  $\sigma$. }
\begin{center}
\begin{tabular}{| c  c |}
\hline
\hline
Parameter & Prior \\  
\hline 
\multicolumn{2}{|c|}{{\bf Cosmology}} \\
$\Omega_m$  &  flat (0.1, 0.9)  \\
$A_s$ &  flat ($5\times 10^{-10},5\times 10^{-9}$)  \\ 
$n_s$ &  flat (0.87, 1.07)  \\ 
$\Omega_b$ &  flat (0.03, 0.07)  \\ 
$h$  &  flat (0.55, 0.91)   \\ 
$\Omega_\nu h^2$  & flat($5\times 10^{-4}$,$10^{-2}$) \\
\hline 
\multicolumn{2}{|c|}{{\bf Lens Galaxy Bias}} \\ 
$b_{i} (i=1,5)$   & flat (0.8, 3.0) \\
\hline
\multicolumn{2}{|c|}{{\bf Intrinsic Alignment}} \\ 
\multicolumn{2}{|c|}{{$A_{\rm IA}(z) = A_{\rm IA} [(1+z)/1.62]^\salpha$}} \\ 
$A_{\rm IA}$   & flat ($-5,5$) \\ 
$\salpha$   & flat ($-5,5$) \\
\hline
\multicolumn{2}{|c|}{{\bf Lens photo-$z$ shift (red sequence)}} \\ 
$\Delta z^1_{\rm l}$  & Gauss ($0.008, 0.007$) \\ 
$\Delta z^2_{\rm l}$  & Gauss ($-0.005, 0.007$) \\ 
$\Delta z^3_{\rm l}$  & Gauss ($0.006, 0.006$) \\ 
$\Delta z^4_{\rm l}$  & Gauss ($0.000, 0.010$) \\ 
$\Delta z^5_{\rm l}$  & Gauss ($0.000, 0.010$) \\
\hline
\multicolumn{2}{|c|}{{\bf Source photo-$z$ shift}} \\ 
$\Delta z^1_{\rm s}$  & Gauss ($-0.001, 0.016$) \\ 
$\Delta z^2_{\rm s}$  & Gauss ($-0.019, 0.013$) \\ 
$\Delta z^3_{\rm s}$  & Gauss ($+0.009, 0.011$) \\ 
$\Delta z^4_{\rm s}$  & Gauss ($-0.018, 0.022$) \\
\hline
\multicolumn{2}{|c|}{{\bf Shear calibration}} \\ 
$m^{i}_{\metacal} (i=1,4)$ & Gauss ($0.012, 0.023$)\\
\hline
\end{tabular}
\end{center}
\label{tab:params}
\end{table}

Note that we did not change any assumptions about the nuisance parameters
relative to our previous analysis applied to $\Lambda$CDM and $w$CDM. It is
possible in principle that extensions (e.g. modified gravity) to these
simplest models warrant more complicated modeling and therefore more nuisance
parameters (e.g.\ adopting more complicated parametrizations of galaxy
bias). To address this possibility, we consider a number of more complicated
parametrizations of the systematic effects (described in Sec.~\ref{sec:valid})
with the aim of determining whether we could misidentify a systematic effect
as evidence for an extension. Our tests, also described in that section,
indicate that constraints on the key extension parameters studied in this
paper are not sensitive to these additional parameters. This justifies our
choice not to modify our fiducial nuisance parametrization described in the
bullet-point list above and used previously in Y1KP.  Future, more precise
data will require revisiting these, in addition to potentially extracting
information about these extensions from the modified behavior of astrophysical
nuisance effects.

\subsection{$\Lambda$CDM extensions}\label{sec:extensions}

We now introduce the four extensions to the simplest $\Lambda$/$w$CDM models
that we study in this paper. The cosmological parameters describing these
extensions, along with priors given to them in our analysis, are given
in Table \ref{tab:ext}.

\subsubsection{Spatial Curvature}
\label{sec:omegak}

Standard slow-roll inflation predicts that spatial curvature is rapidly driven
to zero. In this scenario, the amount of curvature expected today is $\Omega_k\simeq 10^{-4}$,
where the tiny deviation from zero is expected from horizon-scale
perturbations but will be very challenging to measure even with future
cosmological data \cite{Leonard:2016evk}. Departures from near-zero curvature
are however expected in false-vacuum inflation, as well as scenarios that give
rise to bubble collisions \cite{Aslanyan:2015pma, Johnson:2015mma}. With
curvature, and ignoring the radiation density whose contribution is negligible in the late universe, the Hubble parameter generalizes to
\begin{equation}
\frac{H(a)}{H_0} = \left [\Omega_ma^{-3} + (1-\Omega_m-\Omega_k) + \Omega_ka^{-2}
\right ]^{1/2}.
\end{equation}
so that $\Omega_k < 0$ corresponds to spatially positive curvature, and the
opposite sign to the spatially negative case.
In this work, we compare constraints on $\Omega_k$ using DES data alone, as well as
with combinations of subsets of the external data described in \S\ref{sec:extdata}.

We do not modify the standard {\tt HALOFIT}
prescription \cite{Smith:2002dz,Takahashi:2012em} for prediction of the
nonlinear power spectrum for nonzero values of $\Omega_k$.  Simulation
measurements of the nonlinear spectrum for nonzero values of $\Omega_k$ do not
exist to sufficiently validate this regime.  However, it is not an
unreasonable a priori assumption that the nonlinear modification to the power
spectrum is only weakly affected by curvature beyond the primary effect
captured in the linear power spectrum being modified. We do incorporate the
impact of $\Omega_k$ in the evolution of the expansion and growth, which is
properly modeled as part of the linear matter power spectrum that is modified
by {\tt HALOFIT}. We verify that this approximation does not significantly
impact our results by comparing to the case where we restrict our data to
scales that are safely `linear' as described in \S\ref{sec:valid} below.

\subsubsection{Extra relativistic particle species}
\label{sec:sterilenu}

Anisotropies in the CMB are sensitive to the number of relativistic particle
species.  The Standard Model of particle physics predicts that the three
left-handed neutrinos were thermally produced in the early universe and their
abundance can be determined from the measured abundance of photons in the
cosmic microwave background. If the neutrinos decoupled completely from the
electromagnetic plasma before electron-positron annihilation, then the
abundance of the three neutrino species today would be
\be
n = \Neff\times 113 \,{\rm cm}^{-3}
\ee
with $\Neff=3$. In actuality, the neutrinos were slightly coupled during
$e^\pm$ annihilation, so $\Neff=3.046$ in the standard
model \cite{Dicus:1982bz,Dodelson:1992km,Mangano:2005cc}. Values of $\Neff$ larger than this
would point to extra relativistic species. The DES observations are less
sensitive to $\Neff$ than the CMB, because the effect of this
parameter in the DES mainly appears via the change in the epoch of matter-radiation equality.
Nevertheless, DES might constrain some parameters that
are degenerate with $\Neff$ so, at least in principle, adding DES observations
to other data sets might provide tighter constraints.

There are well-motivated reasons for exploring possibilities beyond the
standard scenario. First, the most elegant way to obtain small neutrino masses
is the seesaw model~\cite{GellMann:1980vs}, which typically relies on three
new heavy Standard Model singlets, or {\rm sterile} neutrinos. While these
often are unstable and have very large masses, it is conceivable that sterile
neutrinos are light and stable on cosmological timescales~\cite{
Dodelson:1993je}. Indeed, there are a variety of
experimental anomalies that could be resolved with the introduction of light
sterile neutrinos, and a keV sterile neutrino remains an interesting dark
matter candidate. If one or more light sterile neutrinos do exist, then they
would typically be produced in the early universe via oscillations from the
thermalized active neutrinos with an abundance determined by the mixing
angles. As an example, the LSND/Miniboone
anomaly \cite{Aguilar:2001ty,Aguilar-Arevalo:2018gpe} could be resolved with a
light sterile neutrino thus implying $\Neff\simeq 4$; the mixing angle of the
sterile neutrino would dictate that it would have the same abundance as the 3
active neutrinos. More generally, a wide variety of extensions to the Standard
Model contain light stable particles that would have been produced in the
early Universe~\cite{Essig:2013lka} and impacted the value of $\Neff$.
It is important to note that while the addition of an extra relativistic species would explain some aspects of these observations, it is difficult for such models to accommodate all of the existing neutrino oscillation observations.

In the fiducial model, we are allowing for a single free parameter $\sum
m_\nu$, treating the 3 active neutrinos as degenerate (since they would be
approximately degenerate if they had masses in the range we can probe, $> 0.1$
eV). There is some freedom in how to parametrize the extension of a light
sterile neutrino, however. If we attempt to model the addition of a single
sterile neutrino, then in principle two new parameters must be added.
For example, if the sterile neutrino has the same temperature as the active
neutrinos, then the parameters can be chosen to be $\Neff$, allowed to vary
between 3.046 and 4.046, and $m_s$, the mass of the sterile neutrino. Two
light sterile neutrinos would require two more parameters, etc. However, we
expect that the cosmological signal will be sensitive primarily to the total
neutrino mass density and the number of effective massless species at the time
of decoupling, as captured by $\Neff$, so we use only these two parameters,
$\sum m_\nu$ and $\Neff$. Note that a value of $\Neff$ appreciably different
than $3$ would point to a sterile neutrino or another light degree of freedom.
We give $\Neff$ a flat prior in the range $[3.0, 9.0]$, where the lower
hard bound encodes the guaranteed presence of at least three relativistic
neutrino species.

When varying $N_{\rm eff}$, the fraction of baryonic mass in helium $Y_{p}$ is set by a fitting formula based on the PArthENoPE BBN code \cite{Pisanti2008}. This interpolates a $Y_{p}$ for a given combination of $\Omega_b h^{2}$ and $N_{\rm eff}$. An additional prior of $\Omega_b h^{2} < 0.04$ is applied in the $N_{\rm eff}$ analysis to restrict the interpolation to its valid range.

\subsubsection{Time-varying equation-of-state of dark energy}
\label{sec:de_w0wa}

Given the lack of understanding of the physical mechanism behind the
accelerating universe, it is important to investigate whether the data prefer
models beyond the simplest one, the cosmological constant. In Y1KP, we
investigated the evidence for a constant equation-of-state parameter $w\ne-1$. We found no evidence for $w\ne-1$, with a very tight constraint from the combination of DES Y1, CMB, SNe Ia, and BAO of $w=-1.00^{+0.05}_{-0.04}$. 

We now investigate whether there is evidence for the time-evolution of the equation-of-state $w$.
We consider the phenomenological model that describes dynamical dark energy \cite{Linder_wa}
\begin{equation}
w(a) = w_0 + (1 - a) w_a,
\label{eq:wa}
\end{equation}
where $w_0$ is the equation-of-state today, while $w_a$ is its variation with
scale factor $a$. The $(w_0, w_a)$ parametrization fits many scalar
field and some modified gravity expansion histories up to a sufficiently high
redshift, and has been used extensively in past constraints on dynamical
dark energy.

It is also useful to quote the value of the equation-of-state at the pivot
$w_p\equiv w(a_p)$; this is the scale factor at which the equation-of-state
value and its variation with the scale factor are decorrelated, and where $w(a)$
is best-determined. Rewriting Eq.~(\ref{eq:wa}) as $w(a) = w_p + (a_p - a)
w_a$, the pivot scale factor is
\begin{equation}
  a_p = 1+\frac{\text{C}_{w_0 w_a}}{\text{C}_{w_a w_a}}
  \label{eq:wp}
\end{equation}
where $\mathbf{C}$ is the parameter covariance matrix in the 2D $(w_0, w_a)$ space, obtained by marginalizing the full
$28\times 28$ covariance over the remaining 26 parameters. The corresponding
pivot redshift is of course $z_p=1/a_p-1$.

The linear-theory observable quantities in this model are
straightforwardly computed, as the new parameters affect the background
evolution in a known way, given that the Hubble parameter becomes
\begin{equation}
\frac{H(a)}{H_0} = \left [\Omega_ma^{-3} + (1-\Omega_m)a^{-3 (1 + w_0 + w_a)} e^{-3 w_a (1
    - a)}\right ]^{1/2}.
\end{equation}

To obtain the nonlinear clustering in the $(w_0, w_a)$ model, we assume the
same linear-to-nonlinear mapping as in the $\Lambda$CDM model, except for the
modified expansion rate
$H(z)$ \cite{Francis:2007qa,Casarini:2016ysv,Lawrence:2017ost}. In
particular, we implement the same {\tt HALOFIT}
nonlinear \cite{Smith:2002dz,Takahashi:2012em} prescription as we do in the
fiducial $\Lambda$CDM case. We impose a hard prior $w_0+w_a\leq 0$; models
lying in the forbidden region have a positive equation of state in the early
universe, are typically ruled out by data, and would present additional
challenges in numerical calculations. For the same reason we impose the prior
$w_0<-0.33$.  Note also that in our analysis we do implicitly allow the
``phantom'' models where $w(a)<-1$; while not a feature of the simplest
physical models of dark energy (e.g.\ single-field quintessence), such a
violation of the weak energy condition is in general allowed
\cite{Carroll:2003st}.


\def\arraystretch{1.8}      
\setlength{\tabcolsep}{7pt} 
\begin{table}
\caption{Summary of the extensions to the $\Lambda$CDM model that we study in
  this paper, the parameters that describe these extensions, and the (flat)
  priors given to these parameters. In addition to the priors listed in the
  table, we also impose the prior $w_0+w_a\leq 0$ for dark energy, and
  $2\Sigma_0+1>\mu_0$ for modified gravity.}
\label{tab:ext}
\begin{center}
\begin{tabular}{|| c | c | c ||}
\hline
\hline
$\Lambda$CDM Extension & Parameter & Flat Prior \\ \hline\hline
Curvature                              & $\Omega_k$ &   $[-0.25, 0.25]$ \\ \hline
Number relativistic species            & $\Neff$    &   $[3.0, 7.0]$ \\ \hline
\multirow{2}{*}{Dynamical dark energy} & $w_0$      &   $[-2.0,-0.33]$ \\ \cline{2-3}
                                       & $w_a$      &   $[-3.0,  3.0]$ \\ \hline
\multirow{2}{*}{Modified gravity}      & $\Sigma_0$  &   $[-3.0,  3.0]$ \\ \cline{2-3}
                                       & $\mu_0$    &   $[-3.0, 3.0]$ \\ \hline
\hline
\end{tabular}
\end{center}
\end{table}

\subsubsection{Modified gravity}
\label{sec:sigmamu}
The possibility of deviations from general relativity on cosmological scales
has been motivated by the prospect that an alternative theory of gravity could
offer an explanation for the accelerated expansion of the
Universe.
In the past several years, numerous works constraining modifications to
gravity using cosmological data have been published, including from the Planck
team \cite{Ade:2015rim, Aghanim:2018eyx}, the Kilo Degree
Survey \cite{Joudaki:2016kym}, and the Canada-France-Hawaii Lensing
Survey \cite{Simpson:2012ra}.  Constraints from the Dark Energy Survey Science
Verification data were obtained in \cite{Ferte:2017bpf}.  Recently, stringent
constraints were made on certain alternative theories of
gravity \cite{Lombriser:2015sxa,Lombriser:2016yzn,Sakstein:2017xjx,
Ezquiaga:2017ekz,Boran:2017rdn,Baker:2017hug, Kase:2018iwp} via the
simultaneous observation of gravitational and electromagnetic radiation from a
binary neutron star merger with the Laser Interferometer Gravitational Wave
Observatory (LIGO) \cite{TheLIGOScientific:2017qsa}.



In what follows, we refer to the scalar-perturbed Friedmann-Robertson-Walker line element in the conformal Newtonian gauge: 
\begin{equation}
ds^2 = a^2(\tau) \left[ (1 + 2\Psi) d\tau^2 - (1 - 2\Phi) \delta_{ij} dx_i dx_j \right] \,.
\end{equation}
In general relativity and without anisotropic stresses, $\Psi=\Phi$.
The parametrization of deviations from general relativity studied in this
work is motivated by theoretical descriptions which make use of the quasistatic approximation (see, e.g., \cite{Silvestri:2013ne}). It can be shown that in the regime where linear theory holds and where it is a good approximation to neglect time derivatives of novel degrees of freedom (e.g. extra scalar fields), the behavior of the majority of cosmologically-motivated theories of gravity can be summarized via a free function of time and scale multiplying the Poisson equation, and another which represents the ratio between the potentials $\Phi$ and $\Psi$. Such a parametrization is an effective description of a more complicated set of field equations \cite{Baker:2012zs, Creminelli:2008wc, Baker:2011jy, Battye:2012eu, Gleyzes:2013ooa, Gleyzes:2014qga, Bloomfield:2012ff, Amendola:2013qna, Hojjati:2012ci, Hojjati:2013xqa}, but this approximation has been numerically verified on scales relevant to our present work \cite{Noller:2013wca, Schmidt:2009sg, Zhao:2010qy, Barreira:2014kra, Li:2013tda}.

There are a number of related pairs of functions of time and scale which can be used in a quasistatic parametrization of gravity; we choose the functions $\mu$ and $\Sigma$, defined as
\begin{eqnarray}
\label{eq:constraint1}
k^2 \Psi = & -  4 \pi G a^2 (1+\mu(a)) \rho \delta \,, \\
\label{eq:constraint2}
k^2 (\Psi + \Phi) =  & - 8 \pi G a^2 (1+\Sigma(a))  \rho \delta \,,
\end{eqnarray}
where we are working in Fourier space where $k$ is the wavenumber, and
$\delta$ is the comoving-gauge density perturbation.  This version of the
parametrization was used in \cite{Simpson:2012ra, Ade:2015rim,
Aghanim:2018eyx}, and benefits from the fact that $\Sigma$ parametrizes the
change in the lensing response of massless particles to a given matter field,
while $\mu$ is linked to the change in the matter overdensity
itself. Therefore, weak lensing measurements are primarily sensitive to
$\Sigma$ but also have some smaller degree of sensitivity to $\mu$ via their
tracing of the matter field, whereas galaxy clustering measurements depend
only on $\mu$ and are insensitive to $\Sigma$. We find the DES data
alone are more sensitive to $\Sigma$ than to $\mu$; constraining the latter
requires combining DES with a nonrelativistic tracer of large-scale structure
such as the RSD (e.g.\ \cite{Simpson:2012ra,Ferte:2017bpf}) which we also do
below as part of our combined analysis.

To practically constrain $\mu$ and $\Sigma$, we select a functional form of
\begin{equation}
\mu(z) = \mu_0 \frac{\Omega_{\Lambda}(z)}{\Omega_\Lambda}\, , \, \, \, \, \Sigma(z) = \Sigma_0 \frac{\Omega_{\Lambda}(z)}{\Omega_\Lambda}
\label{eq:musigform}
\end{equation}
where $\Omega_{\Lambda}(z)$ is the redshift-dependent dark energy density (in
the $\Lambda$CDM model) relative to critical density, and $\Omega_{\Lambda}$
is its value today.  This time dependence has been introduced
in \cite{Ferreira:2010sz}, and is widely employed (see
e.g. \cite{Simpson:2012ra, Ade:2015rim, Aghanim:2018eyx}). It is motivated by
the fact that in order for modifications to GR to offer an explanation for the
accelerated expansion of the Universe, we would expect such modifications to
become significant at the same timescale as the acceleration begins. We do not
model any scale-dependence of $\mu$ and $\Sigma$ since it has been
shown to be poorly constrained by current cosmological data while not
much improving the goodness-of-fit \cite{Ade:2015rim}. We
therefore include only the parameters $\mu_0$ and $\Sigma_0$ (but, as
explained in Sec.~\ref{sec:valid_sim}, only quote constraints on $\Sigma_0$).
In GR, $\mu_0=\Sigma_0=0$.

Note that although our choice of parametrization is motivated by the
quasistatic limit of particular theories of gravity, our analysis takes an
approach which is completely divorced from any given theory. We endeavor
instead to make empirical constraints on the parameters $\mu_0$ and $\Sigma_0$
as specified by Eqs.~(\ref{eq:constraint1}), (\ref{eq:constraint2}),
and (\ref{eq:musigform}). Because we take this empirically driven approach, we
include certain data elements in which the quasistatic approximation would not
be expected to hold, most importantly the near-horizon scales for the ISW
effect. Although not rigorously theoretically justified, a similar approach
with respect to inclusion of the ISW effect at large scales was taken in, for
example, \cite{Simpson:2012ra}. Practically, this choice has the benefit of
providing an important constraint on $\tau$ from external CMB data, which is
useful in breaking degeneracies.

We use \cosmosis\ with a version of {\tt MGCamb}\footnote{{\tt
https://aliojjati.github.io/MGCAMB/mgcamb.html}} \cite{Zhao_2009,
Hojjati:2011ix} modified to include the $\Sigma$, $\mu$ parametrization to
compute the linear matter power spectrum and the CMB angular power spectra.
For some sets of ($\Sigma_0$, $\mu_0$) {\tt MGCamb} returns an error; we
estimated this region of parameter space can be avoided by imposing an
additional hard prior $\mu_0<1+2\Sigma_0$. We therefore implement this prior
in order to avoid computations for parameters not handled by {\tt
MGCamb}. The effects of this hard prior can be visually observed in our
constraints on modified-gravity parameters with DES alone (Fig.~\ref{fig:w0wa_MG}
below); we demonstrate in Appendix  \ref{app:mg_cosmosislike} that its effects on the combined DES+external
constraint is likely to be minimal.

To validate our modified-gravity analysis pipeline, we compare the \cosmosis\
results to that of another code, \cosmolike~\cite{Krause:2016jvl}. We require
that the two codes give the same theory predictions for clustering and lensing
observables, and the same constraints on cosmological parameters given a
synthetic data vector.  The comparison shows good agreement, and details can
be found in Appendix \ref{app:mg_cosmosislike}.

Finally, because the $(\mu, \Sigma)$ description does not constitute a
complete theoretical model, its nonlinear clustering predictions are not
available to us even in principle. We therefore restrict ourselves to the
linear-only analysis. To do this, we follow the Planck 2015
analysis \cite{Ade:2015rim} and consider the difference between the nonlinear
and linear-theory predictions in the standard $\Lambda$CDM model at best-fit
values of cosmological parameters and with no modified gravity. Using the
respective data vector theory predictions, $\mathbf{d}_{\rm NL}$ and
$ \mathbf{d}_{\rm lin}$, and full error covariance of DES Y1, $ \mathbf{C}$, we
calculate the quantity
\begin{equation}
\Delta\chi^2 \equiv (\mathbf{d}_{\rm NL}-\mathbf{d}_{\rm lin})^T\,
\mathbf{C}^{-1}\,(\mathbf{d}_{\rm NL}-\mathbf{d}_{\rm lin})
\end{equation}
and identify the single data point that contributes most to this quantity. We
remove that data point, and repeat the process until $\Delta\chi^2<1$. The
resulting set of 334 (compared to the original 457) data points that remain
constitutes our fiducial choice of linear-only scales.

\Swide{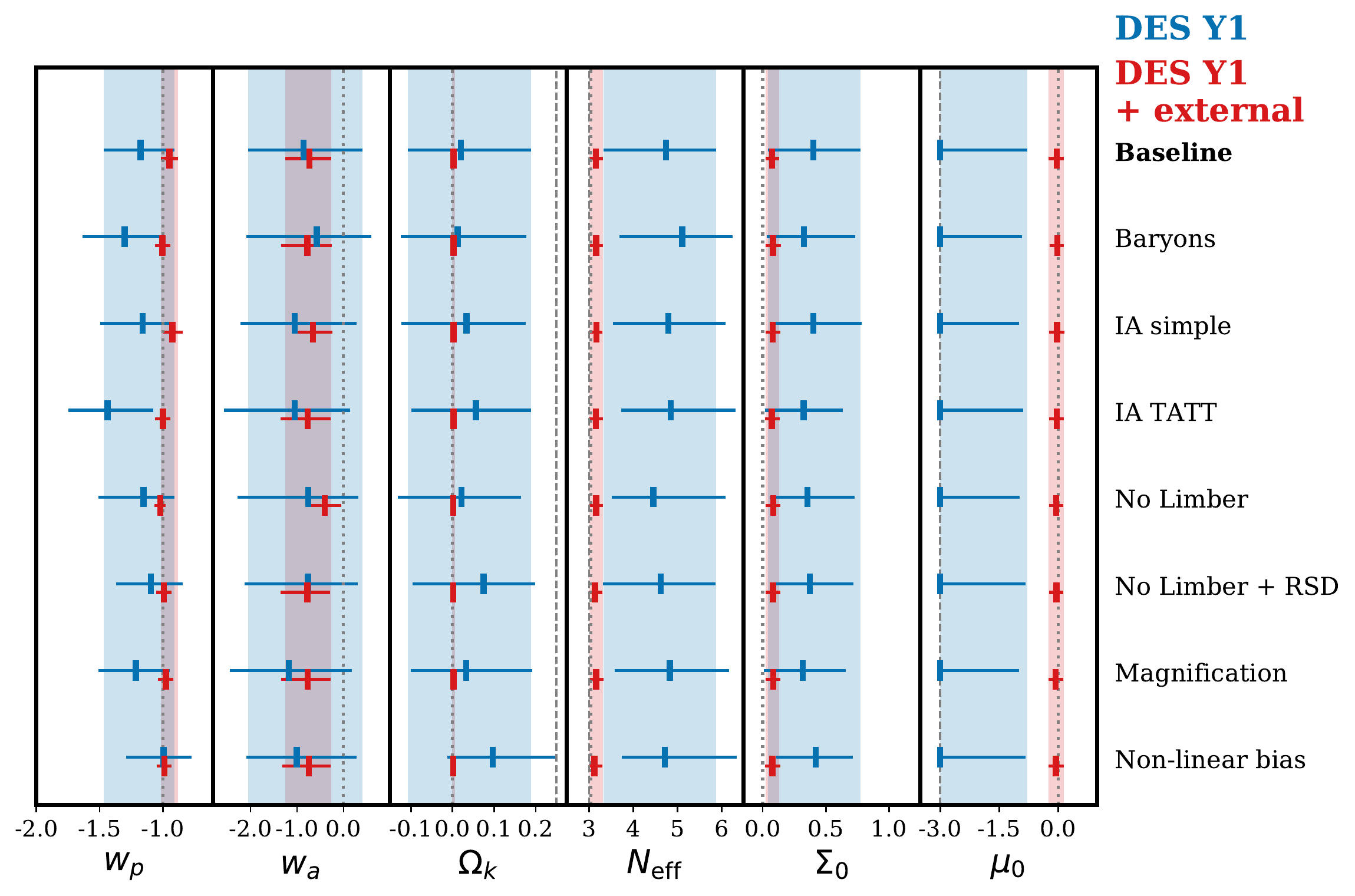}{Impact of assumptions and approximations
  adopted in our analysis, demonstrated on synthetic data (that is, noiseless
    DES data centered on the theoretical expectation, along with actual external
  data).  Each column shows one of the cosmological parameters describing
  $\Lambda$CDM extensions;  the dotted vertical line is the true input
  value of that parameter in the DES data vector (which does not necessarily
  coincide with the parameter values preferred by the external data). The
  vertical shaded bands show the marginalized 68\% CL constraints in the baseline
  model for the DES-only synthetic data (blue) and DES+external. The
  horizontal error bars show the inferred constraint for each individual
  addition to the synthetic data vector which are listed in rows; they match
  the shaded bands for the baseline case. For subsequent rows, they show the
  inferred constraint for each individual addition to the synthetic data
  vector as listed on the right. Some cases that appear inconsistent with the
  baseline analysis are discussed further in Sec.~\ref{sec:valid_sim}. In cases 
  where the prior is informative, we also include a dashed vertical line to signify the prior edge.}

\section{Validation tests and blinding}
\label{sec:valid}

We subject our $\Lambda$CDM extensions analyses to the same battery of tests
for the impact of systematics as in Y1KP. The principal goal is to ensure that
all of our analyses are robust with respect to the effect of reasonable
extensions to models of astrophysical systematics and approximations in our
modeling.  As part of the same battery of tests, we also test that the range
of spatial scales that are used lead to unbiased cosmological results, and
that motivated modifications to our modeling assumptions do not significantly
change the inferred cosmology.

In these tests and the results below,\footnote{One important distinction from
  the data-based results in later sections is that we sample a lower-precision
  version of the CMB lensing contribution to constraints including external
  data when varying $\Omega_k$, then modify the posterior to the higher
  precision prediction via importance sampling. We do this to speed up
    the evaluation of nonflat models, as in our implementation of CAMB,
    particularly when evaluating CMB lensing, sampling over many
    chains at full precision is impractical. We checked that this
    approximation has a minor effect on the shape
    of the posterior.}  sampling of the posterior distribution of the
  parameter space is performed with {\tt Multinest} \cite{Feroz:2008xx} and
  {\tt emcee} \cite{ForemanMackey:2012ig} wrappers within
  \cosmosis\footnote{{\tt https://bitbucket.org/joezuntz/cosmosis/}}
  \cite{Zuntz:2014csq} and \cosmolike ~\cite{Krause:2016jvl}. While the
  convergence of {\tt Multinest} is intrinsic to the sampler and achieved by
  verifying that the uncertainty in the Bayesian evidence is below than some
  desired tolerance, we explicitly check the convergence of {\tt emcee}
  chains. In order to do so, we compute the autocorrelation length of each
  walk, then continue the walks until a large number of such lengths is
  reached\footnote{The recommended methods for convergence testing (as well as
    the documentation for {\tt emcee}) can be found in {\tt
      https://emcee.readthedocs.io/}}. The autocorrelation length estimates
  how long a chain needs to be in order for new ``steps'' to be uncorrelated
  with previous ones. We then split chains into several uncorrelated segments
  and verify that marginalized parameter constraints do not change
  significantly when these segments are compared with each other. The typical
  number of samples of the posterior in these chains is between two and three
  million. We have also verified in select cases that this procedure leads to
  excellent agreement with the 1D marginalized parameter posteriors achieved
  by {\tt Multinest}, so both samplers are used interchangeably in what
  follows.

\subsection{Validation of assumptions using synthetic data}\label{sec:valid_sim}

In order to verify that our results are robust to modeling assumptions and
approximations, we compare the inferred values of the extension parameters
($\Omega_k,N_{\rm eff},\ldots$) obtained by a systematically shifted,
noiseless synthetic data vector. The synthetic data vector is
centered precisely on  the standard $\Lambda$CDM
cosmology, except it is shifted with the addition of a
systematic effect that is {\bf not} included in our analysis.  The goal is to
ensure that we do not claim evidence for an extension to $\Lambda$CDM when the
real data contains astrophysical effects more complex than those in our model.
For each systematic effect, we compare the inferred set of extension
parameters to the fiducial, unmodified extension parameters used to create the synthetic data (which we refer to as the ``baseline'' constraint). For all of
these tests, for DES we use the synthetic data vectors (for the baseline case
and the systematic shifts described below), but for the external data sets ---
CMB, BAO, RSD, and SN Ia --- we use the actual, observed data vector.

The changes to modeling assumptions that we consider are:
\begin{enumerate}
\item Baryonic effects: we synthesize a noiseless data vector including a contribution to
  the nonlinear power spectrum caused by AGN feedback using the OWLS AGN
  hydrodynamical simulation \cite{owls_agn} and following the methodology of
  \cite{shearcorr}.

\item Intrinsic alignments, simple case: we synthesize a noiseless data vector with the IA
  amplitude $A_{\rm IA} = 0.5$ and redshift scaling $\salpha = 0.5$ using the
  baseline nonlinear alignment model used in Y1KP. While we explicitly
  marginalize over these IA parameters in our analysis, this systematic check
  is still useful to monitor any potential biases due to degeneracy between
  the cosmological parameters and $(A_{\rm IA},\salpha$) and the
    presence of non-Gaussian posteriors.

\item Intrinsic alignments, complex case: we synthesize a
  noiseless data vector using a subset of the tidal alignment and tidal
  torquing model (hereafter TATT) from \cite{Blazek17}. This introduces a
  tidal torquing term to the IA spectrum that is quadratic in the tidal
  field. The TATT amplitudes were set to $A_{1}=0$, $A_{2}=2$ with no $z$
  dependence, as was done in \cite{methodpaper} when validating the analysis
  of Y1KP.

\item Nonlinear bias: we test our fiducial linear-bias assumption by
  synthesizing a noiseless data vector that models the density contrast of galaxies as
\begin{equation}
\delta_g = b_1^i \delta + \frac{1}{2} b_2^i [\delta^2 - \sigma^2]
\end{equation}
where $\delta$ and $\delta_g$ are the overdensities in matter and galaxy
counts respectively, and the density variance $\sigma^2 = \langle
  \delta^2 \rangle$ is subtracted to enforce $\langle \delta_g \rangle = 0$.
  While this relationship is formally defined for smoothed density fields, the
  results do not depend on the choice of smoothing scale since, e.g., the
  variance explicitly cancels with contributions to the two-point correlation.
  We are considering scales that are sufficiently larger than the typical
  region of halo formation that we neglect higher-derivative bias terms.  See
  Ref.~\cite{Desjacques:2016bnm} for further discussion of nonlinear biasing.
Here $i$ refers to the lens redshift bin and where $b_1^i=\{1.45, 1.55, 1.65,
1.8, 2.0\}$ for the five bins. The $b_2$ values used for each lens bin were
estimated from from the following relationship fit in simulations
  \cite{Lazeyras:2015lgp}: $b_2 = 0.412 - 2.143\,b_1 + 0.929\,b_1^2 +
0.008\,b_1^3$.  Because the contribution from tidal bias $b_{s^2}$ is
  expected to be small, we set it to zero in these validation tests.

\item Magnification: we synthesize a noiseless data vector that includes the contribution
  from magnification to $\gamma_{t}$ and $w(\theta)$. These are added in
  Fourier space using \cite{Bernstein:2008aq}.

\item Limber approximation and RSD: we synthesize a noiseless data vector that uses the
  exact (non-Limber) $w(\theta)$ calculation\footnote{We do
      not investigate the effect of the Limber approximation on the tangential
      shear profile $\gamma_{t}$, since it includes the projection from the
      observer to the source galaxy, and is less sensitive to the Limber
      approximation, below the level of the DES Y1 statistical uncertainty
      \cite{dePutter:2010jz}.} and include the contribution from redshift
    space distortions \cite{Padmanabhan07}.
    
\end{enumerate}
More information about the implementation of these tests can be found in \cite{methodpaper}.

The results of these tests are shown in
Fig.~\ref{fig:tableplot_extensions_systests}. The columns show the parameters
describing $\Lambda$CDM extensions, namely $w_p$, $w_a$, $\Omega_k$, $\Neff$,
$\Sigma_0$, and $\mu_0$. The shaded vertical region shows the marginalized
68\% posterior confidence limit (CL) in each parameter for the baseline case.
The horizontal error bars show how this 
posterior, fully marginalized over all other parameters, including the
  other parameter in two-parameter extensions,  changes with the systematic
described in the given row for the case of DES-only (blue bars) and
DES+external (red bars) data. We observe that, except in the cases explained
below, the marginalized posteriors are consistent with the baseline analysis
in these tests\footnote{The DES+external $\Omega_k$ column in
    Fig.~\ref{fig:tableplot_extensions_systests} is narrow and hard to
    visually inspect, but we have verified that there are no biases in the
    curvature parameter with alternate assumptions about the systematic errors
    shown shown in the different rows.}.

Figure \ref{fig:tableplot_extensions_systests} shows shifts in some DES-only
68\% C.L. constraints relative to the input value shown by the dotted vertical
lines. The most pronounced effect is in the DES-only case for modified gravity
parameter $\mu_0$ (and, to a slightly smaller extent, $\Sigma_0$ and $\Neff$), which is more than
1-$\sigma$ away from its true value of zero. Upon investigating this, we found
that the bias away from the input value is caused by the interplay of two
effects: 1) weak constraints, with a relatively flat likelihood profile in
these parameters in certain directions, combined with 2) prior-volume effect,
where the large full-parameter-space volume allowed in the direction in which
the parameter is a reasonably good fit ends up dominating the total integrated
posterior, resulting in a 1D marginalized posterior that is skewed away from
the maximum likelihood true value. For example, with the restricted range of
scales that we use for the modified gravity tests, negative values of $\mu_0$
are an acceptable (though not the best) fit and, because of the relatively
large number of combinations of other parameters that result in a good
likelihood for $-3\leq \mu_0\lesssim 0$, the 68\% C.L. constraint on $\mu_0$
ends up excluding the input best-fit value of zero (see
Fig.~\ref{fig:tableplot_extensions_systests}). We have explicitly checked that
removing the principal degeneracy with other parameters --- in modified
gravity tests, achieved by fixing the bias parameters $b_i$ --- removes the
bias in $\mu_0$. Nevertheless, because these tests imply that the DES-only
constraint on this parameter would suffer from the aforementioned bias,
we choose not to quote constraints on $\mu_0$ from the DES-only data in the results below.

We also observe a bias in the DES+external constraint on $w_a$ relative to the
input value of zero. This is mostly driven by the fact that the best fit of
the external data does not necessarily coincide with the cosmological
parameter values assumed for the synthetic data vectors used to produce DES
constraints -- in fact, it is well-known that external data alone favor
$w_a<0$ \cite{Aghanim:2018eyx}. Additionally, even the DES synthesized data alone mildly prefer
negative $w_a$ due to the prior-volume effect mentioned above. The resulting synthesized
DES+external constraint on $w_a$ is then biased negative at greater than 68\%
confidence. Because the combined analysis on the real data will not be subject
to the principal cause of the $w_a$ bias observed here, we proceed with the
analysis.

There are therefore two takeaways from
Fig.~\ref{fig:tableplot_extensions_systests}:
\begin{itemize}
  \item First, the projected 1D inferences from DES-only measurements on
    $\mu_0$ are likely to be biased principally due to the prior volume
    effect, so we choose not to quote constraints on this parameter in the
    DES-only case (but still include it in the analysis throughout).  We do not attempt
    to correct the biases in the $w_0$-$w_a$ DES+external case or inflate the
    parameter errors to account for it; see the discussion above.
\item Second and most importantly, the different assumptions considered in
  Fig.~\ref{fig:tableplot_extensions_systests} produce consistent results with
  the baseline constraint for all parameters describing $\Lambda$CDM
  extensions.
\end{itemize}

\subsection{Validation of assumptions using DES data}\label{sec:valid_analysis}

In addition to the tests in the previous section that constrain potential
biases due to our modeling assumptions and approximations on
synthesized noiseless data, we implement several
validation tests that modify how we analyze the actual DES data vector. In
particular, we test the following assumptions:

\begin{enumerate}[resume]  
\item Intrinsic alignments, free redshift evolution: while the fiducial analysis
  assumes IA to scale as a power-law in redshift (see
  Sec.~\ref{sec:nuisance}), we relax that here by assuming four uncorrelated
  constant amplitudes per source redshift bin.
\item Conservative scales: to gauge how our results depend on the range of angular scales
  used, we adopt the conservative set of (basically linear) scales used in the
  modified gravity extension, and apply it to the other three extensions
  (curvature, $\Neff$, dynamical dark energy).
\item Alternate photometric redshifts: to investigate the robustness of our
  results to the shape of the redshift distribution of source galaxies, we adopt the
  distributions obtained directly from resampling the COSMOS data, as described in \cite{photoz}.
\end{enumerate}

For each of these alternate analysis options, we investigate how the fiducial
constraints on the $\Lambda$CDM extensions parameters change. These results
are presented and discussed along with our main results, near the end of
Sec.~\ref{sec:results}. 

\begin{figure*}[t] 
\centering
\includegraphics[width=0.47\textwidth]{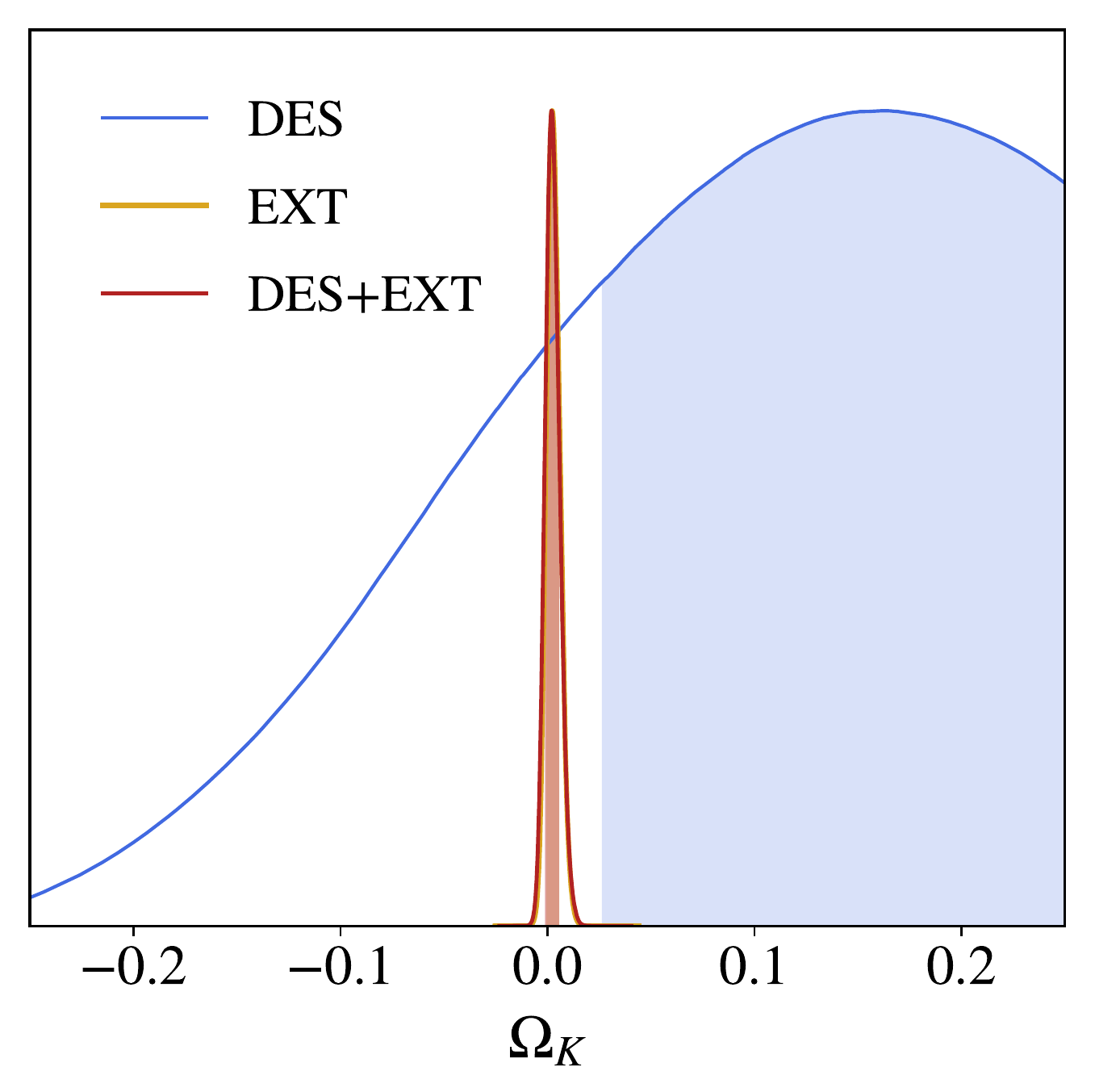}
\includegraphics[width=0.47\textwidth]{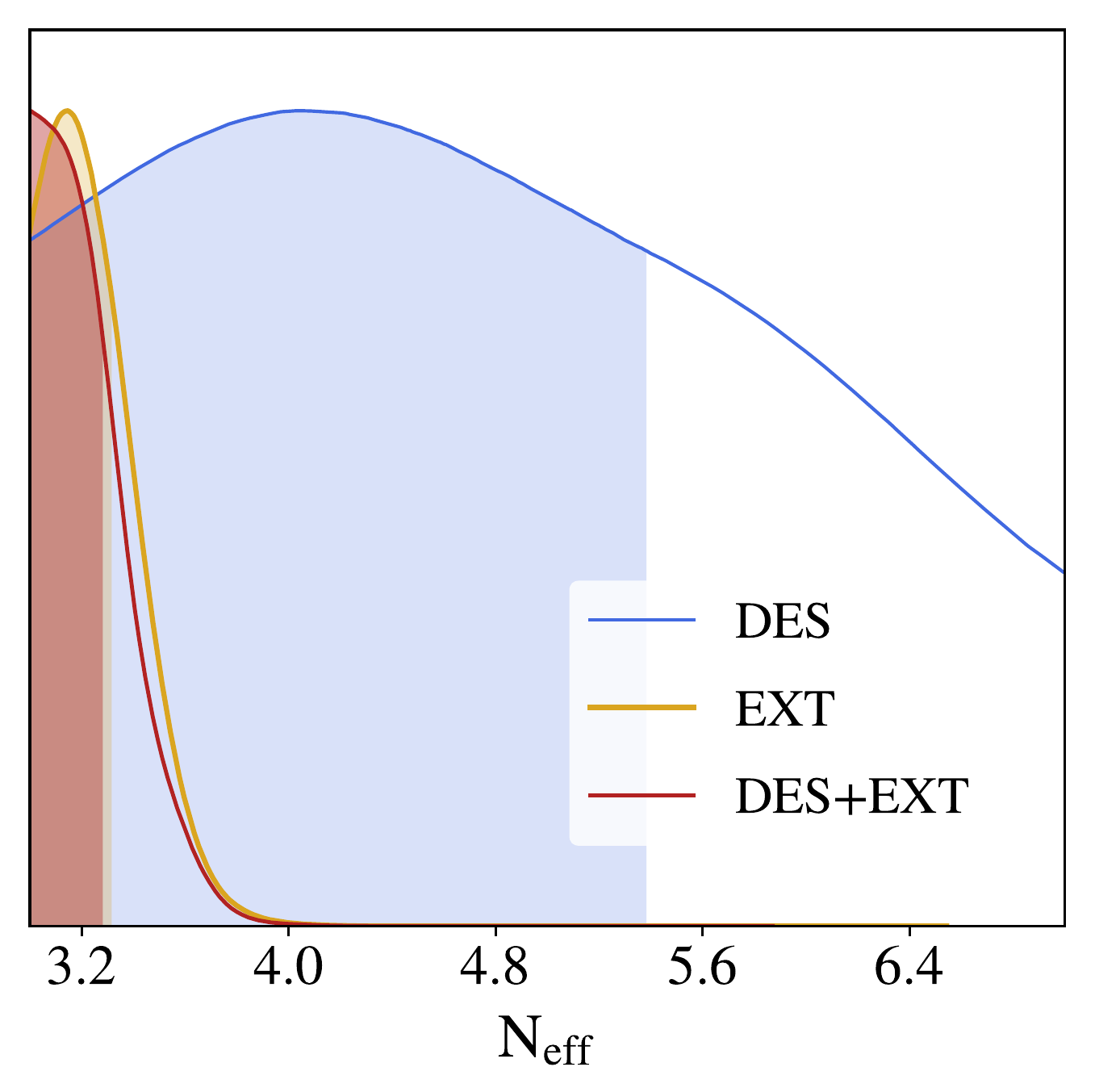}
\caption{Posterior constraints on the spatial curvature (left panel) and the
  number of relativistic species (right panel) in two of the extensions to
  $\Lambda$CDM considered in this paper. Blue contours show DES alone, yellow
  is external data alone, and red is the combination of the two. The 68\%
  confidence region is shaded. The x-axis ranges in both panels coincide with
    the priors given to $\Omega_k$ and $\Neff$, respectively. Posteriors' maxima are
  normalized to unity for better visibility of the DES only results.}
\label{fig:ok_Neff}
\end{figure*}

\subsection{Blinding}

We follow the same strategy as in Y1KP, and blind the principal cosmological
results to protect against human bias. We do so by shifting axes in all plots
showing the cosmological parameter constraints. Where relevant, this includes
simultaneously not plotting theory predictions (including simulation outputs
as ``theory'') in those same plots.  A different shift was applied to each of
the DES, external data, and joint constraint contours in any figures made at
the blinded stage. Moreover evidence ratios of the joint constraints were not
read before unblinding. This was done to prevent confirmation bias based
on the level of agreement between the DES and external constraints.

We unblinded once we ensured that there are no biases on the extension
parameters due to systematics, as shown in
Figs.~\ref{fig:tableplot_extensions_systests} and
\ref{fig:tableplot_extensions_pipelineval}, apart from those that have a
known, statistical explanation (see Sec.~\ref{sec:valid_sim}).  

We have made two modifications to the analysis after the results were
unblinded. First, we identified that the incorrect Planck data file ({\sc
  plik\_lite\_v18\_TTTEEE.clik}) was used for our $(w_0,w_a)$ results and
reran these chains with the correct file ({\sc plik\_lite\_v18\_TT.clik}).  We
verified that this modification does not lead to appreciable differences in
the final constraints, though it does lead to a difference in the reported
Bayesian evidence ratios for this case. Second, we adopted the GetDist code to
evaluate the marginalized posteriors, as it is more suitable to handle
boundary effects in the posteriors~\cite{camb_notes}. This leads to small
differences in cases where the constraints are strongly informed by the prior
boundaries, such as $\Neff$.

\section{Results}
\label{sec:results}

The constraints on curvature and the number of relativistic species are
given in the two panels of Fig.~\ref{fig:ok_Neff}. For curvature, we find

\begin{equation}
  \begin{aligned}  
 \Omega_k &= 0.16^{+0.09}_{-0.14}\qquad \text{DES\ Y1}\\[0.2cm] 
          &= 0.0020^{+0.0037}_{-0.0032}\quad \text{DES\ Y1 + External}
 \end{aligned}
\end{equation}
while for the number of relativistic species, the lower limit hits against our
hard prior of $\Neff>3.0$ so we quote only the 68\% (95\%) upper limits
\begin{equation}
  \begin{aligned} 
 \Neff    &< 5.28 \,(\mbox{---})    \qquad \text{DES\ Y1}\\[0.2cm]
          &< 3.28 \,(3.55)   \quad \text{DES\ Y1 + External}.
 \end{aligned}
\end{equation}
where the dashes indicate that we do not get a meaningful upper limit from DES
alone at the 95\% since the constraint hits against the upper limit of our
prior.

Figure~\ref{fig:ok_Neff} indicates that DES alone constrains curvature weakly,
showing mild ($\sim 1$-$\sigma$) preference for positive values of $\Omega_k$; note
also that this constraint is informed by the upper prior boundary. The
DES-only constraint on $\Neff$ is also relatively weak, and is fully
consistent with the theoretically favored value $\Neff=3.046$. Moreover, 
the DES Y1 data do not appreciably change the existing external-data
constraints on these two parameters. The addition of the DES data to external
measurement does slightly suppress $\Neff$, which can be understood as
follows.  The DES data prefer a lower $\Omega_m$ than the external data,
leading to a slight increase in $h$ such that the posterior distribution in
$\Omega_m h^3$ is downweighted at the high values of this parameter
combination. Because $\Omega_m h^3$ is highly correlated with $\Neff$ --- they
both generate out-of-phase changes in the CMB temperature power spectrum ---
adding DES to external data also has the consequence of slightly suppressing
$\Neff$.

We also compare the cases where the number of relativistic species is fixed at
$\Neff=3.046$ (the standard model) and $\Neff=4.046$ (standard model, plus a
single fully thermalized sterile neutrino). Preference for one model over the
other is assessed using the evidence ratio,
\begin{equation}
R^{\Neff} = \frac{P(\mathbf{d}|\Neff=4.046)}{P(\mathbf{d}|\Neff=3.046)}. 
\end{equation}
where $P(\mathbf{d}|\Neff)$ is the Bayesian evidence, given by the integral over the
parameter space of the likelihood times the prior; see Eq.~(5.1) in Y1KP.
A ratio much greater than 1 would imply $\Neff=4.046$ is favored and a ratio
much less than 1 would imply that $\Neff=3.046$ is favored.  The Bayesian
evidence ratios for DES alone is $R^{\Neff}=0.78$, indicating no statistical
preference for an extra relativistic species. For the external data alone and
DES plus external data, the ratios are $R^{\Neff} = 0.0033$ and $R^{\Neff} =
0.0049$, respectively. The combined data therefore show strong evidence to
support the standard value $\Neff=3.046$ relative to the case with one
additional relativistic species; DES does not appreciably change the result
obtained using the external data alone (the apparent increase on the odds of
$\Neff=4$ when going from external to DES+external data is not statistically
significant as the errors on $R$ are larger than the difference between these
two values.)

\begin{figure*}[] 
\centering
\includegraphics[width=0.47\textwidth]{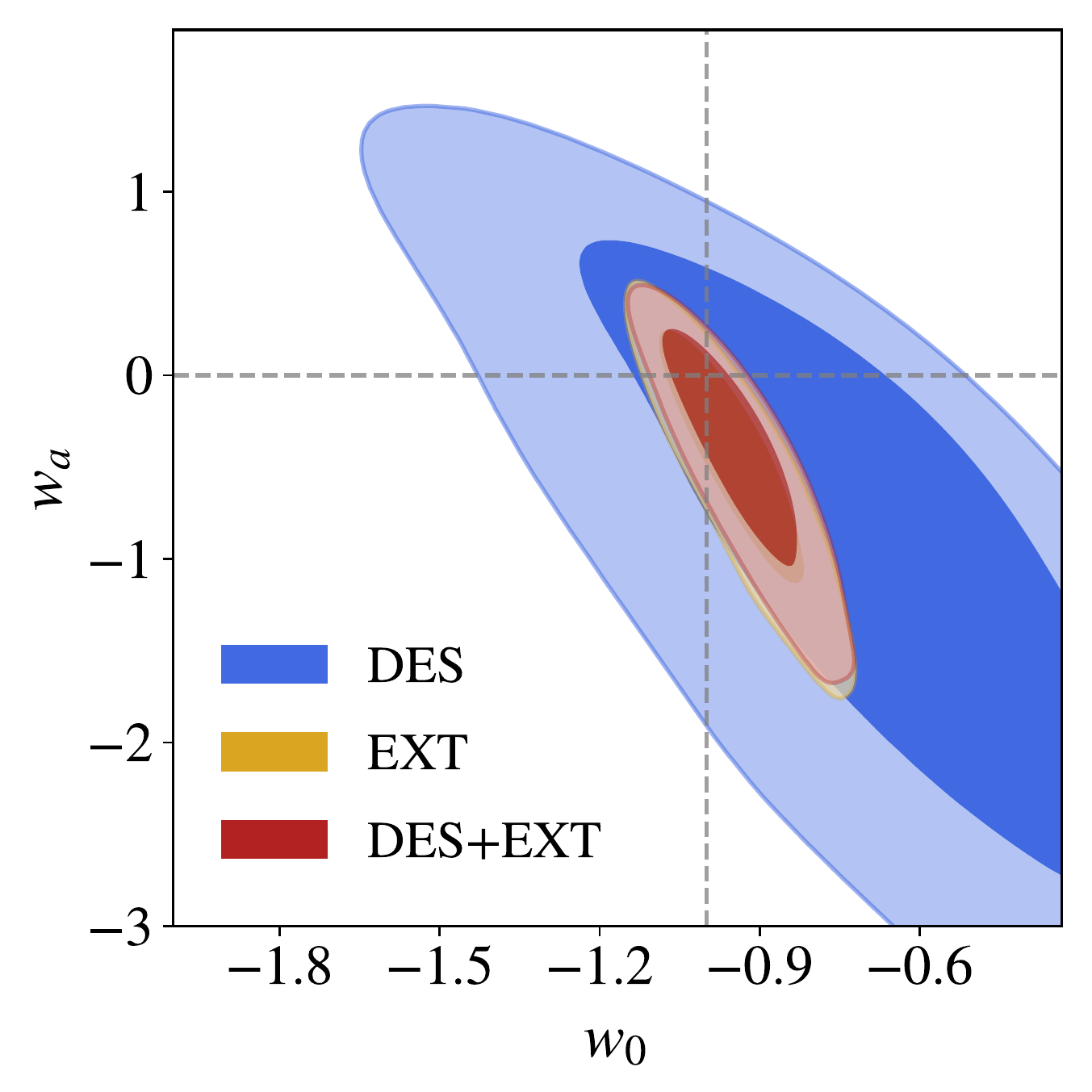}
\includegraphics[width=0.47\textwidth]{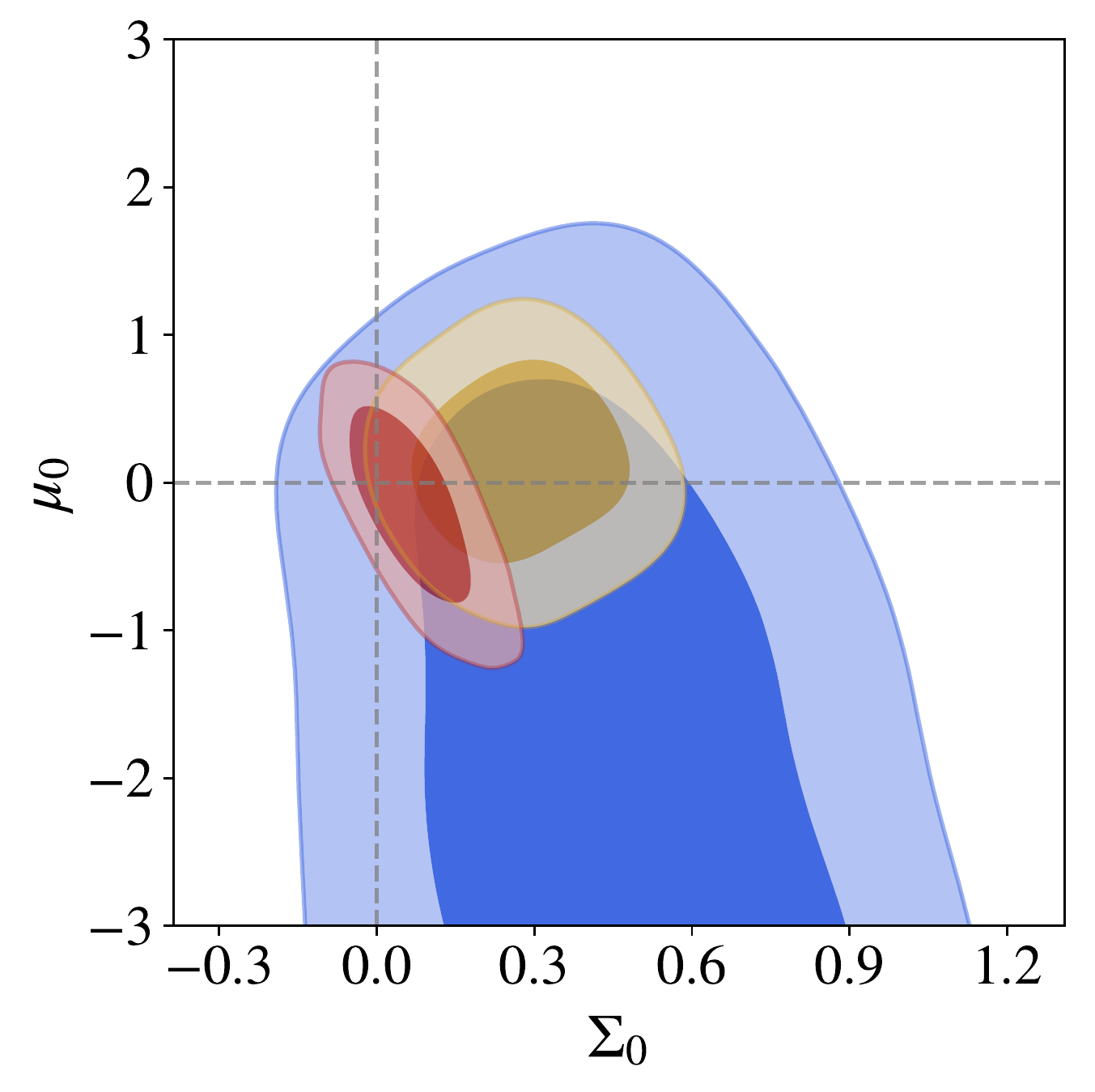}
\caption{Constraints on dark energy parameters $(w_0, w_a)$ (left panel) and
  the modified gravity parameters $(\Sigma_0, \mu_0)$ (right panel). Blue
  contours show the 68\% and 95\% confidence regions from DES alone, yellow is
  external data alone, and red is the combination of the two. The intersection
  of the horizontal and vertical dashed lines shows the parameter values in
  the $\Lambda$CDM model (left panel) and in general relativity (right).
  The x-axis range in the left panel and the y-axis range in the right
    panel coincide with the respective priors given to $w_0$ and $\mu_0$. The
  cause of the nonintuitive shift in the combined $\Sigma_0$ constraint (red
  contour) relative to separate constraints is discussed in
  Sec. \ref{sec:results}.  }
\label{fig:w0wa_MG}
\end{figure*}

\begin{figure}[] 
\centering
\includegraphics[width=0.47\textwidth]{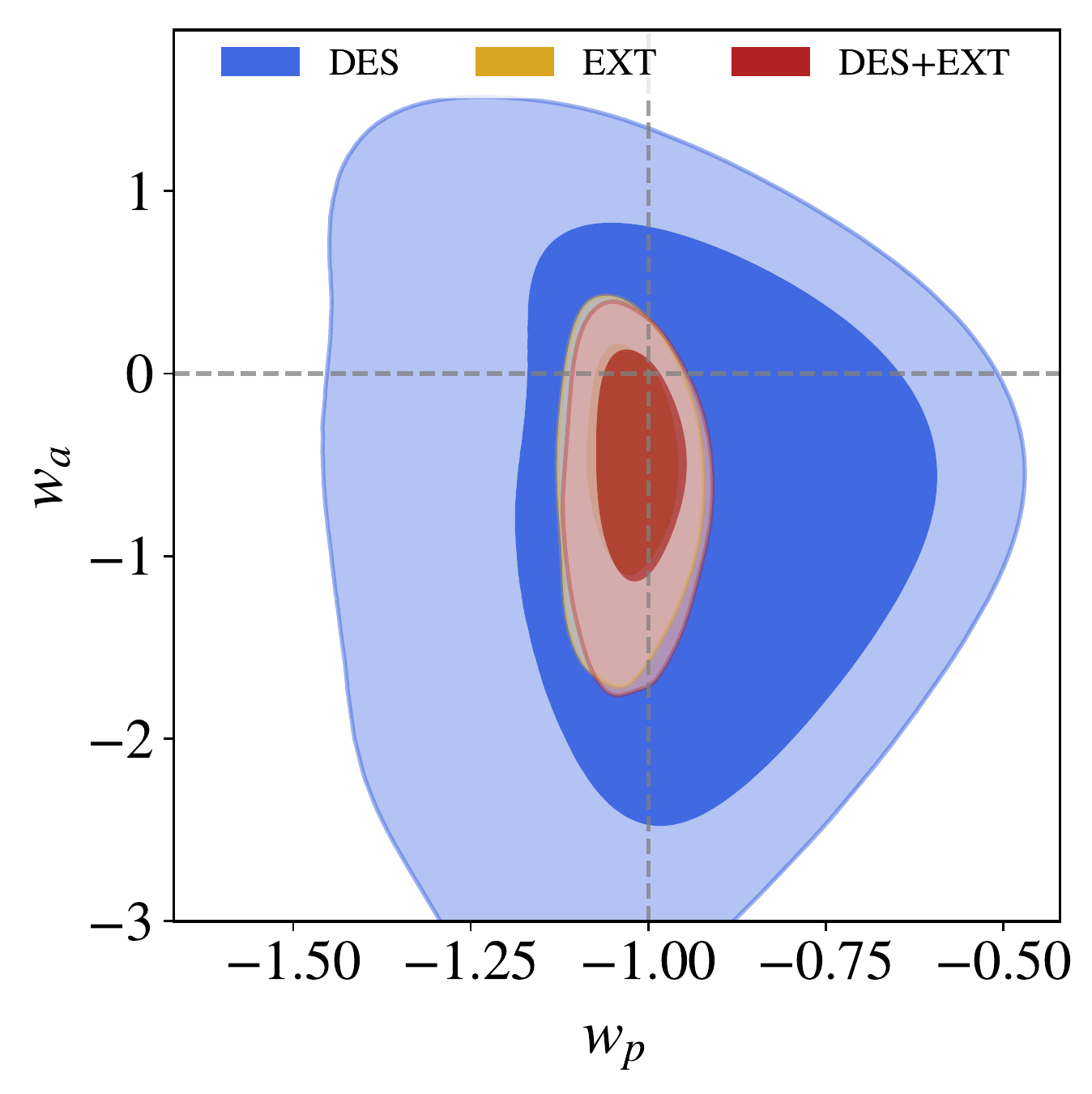}
\caption{Constraints on the pivot value of the dark energy equation-of-state
  $w_p$ and the variation with scale factor $w_a$ Blue contours show DES
  alone, yellow is external data alone, and red is the combination of the two.
  The intersection of the horizontal and vertical dashed lines shows the
  parameter values in the $\Lambda$CDM model.}
\label{fig:wp_wa}
\end{figure}

We now turn to dynamical dark energy. The constraints are shown in the left
panel of Fig.~\ref{fig:w0wa_MG}. We find
\begin{align*} 
 w_0 &= -0.69^{+0.30}_{-0.29},\;\;& w_a &= -0.57^{+0.93}_{-1.11}\qquad \text{DES\ Y1}\\[0.2cm]
     &= -0.95^{+0.09}_{-0.08},\;\;&     &= -0.28^{+0.37}_{-0.48}\qquad \text{DES\ Y1 + Ext}.
\end{align*}
The DES Y1 data alone are therefore consistent with the cosmological-constant values
of $(w_0, w_a)=(-1, 0)$; they do not appreciably change the constraint from
external data alone.

The pivot equation-of-state (see definition in Eq.~(\ref{eq:wp})) is obtained to be
\begin{equation}
  \begin{aligned} 
 w_p     &= -0.91^{+0.19}_{-0.23}\qquad \text{DES\ Y1}\\[0.2cm]
         &= -1.01^{+0.04}_{-0.04}\qquad \text{DES\ Y1 + External}.
 \end{aligned}
\end{equation}
For the DES-only and DES + External cases, the pivot redshift is found to be
$z_p = 0.27$ and $z_p = 0.20$, respectively. Figure ~\ref{fig:wp_wa} shows the
constraints in the $(w_p, w_a)$ plane.

Do the DES data favor the introduction of two new parameters, $w_0$ and $w_a$, to the
$\Lambda$CDM model? Again, we calculate the Bayesian evidence ratio 
\begin{equation}
R^{(w_0, w_a)} = \frac{P(\mathbf{d}|w_0, w_a)}{P(\mathbf{d}|w_0=-1, w_a=0)}. 
\end{equation}
For DES data alone, we find $R^{(w_0, w_a)} = 0.11$, while the DES+external
data give $R^{(w_0, w_a)} = 0.006$. Therefore, Bayesian evidence ratios strongly
support $\Lambda$CDM, and do not favor introduction of the additional parameters
$w_0$ and $w_a$.

Finally, we turn our attention to modified gravity, the extension for which DES carries the most weight. Recall from
Sec.~\ref{sec:valid_sim} that we have decided to quote only the constraint on the
parameter $\Sigma_0$ in the DES-only case. The constraint, shown in the right panel of
Fig.~\ref{fig:w0wa_MG}, is
\begin{equation}
  \begin{aligned} 
    \Sigma_0 &= 0.43^{+0.28}_{-0.29}  & 
     \text{DES\ Y1}\\[0.2cm]
    \Sigma_0 &= 0.06^{+0.08}_{-0.07},\quad \mu_0 = -0.11^{+0.42}_{-0.46} & \text{DES\ Y1 + Ext},
  \end{aligned}
\end{equation}
the latter of which can be compared to the external-only constraint, which is
$\Sigma_0=0.28^{+0.13}_{-0.14}$.  Thus the addition of DES data improves the
constraints on $\Sigma_0$ by almost a factor of two.
\def\arraystretch{1.8}      
\setlength{\tabcolsep}{7pt} 
\begin{table*}
\caption{Constraints on the parameters describing the extensions of the
  $\Lambda$CDM model that we study in this paper. All errors are 68\%
  confidence intervals, except for $\Neff$ where we show the 68\% upper bound.
  We do not quote the DES-only constraint on $\mu_0$, as discussed in
  Sec.~\ref{sec:valid_sim}.  The last column shows the improvement in the
    goodness-of-fit, $\Delta\chi^2$, between the corresponding best-fit
    extension and the best-fit $\Lambda$CDM.  Note that the sampling error in
    the $\Delta\chi^2$ values is $\sim$0.5; hence, the two positive values in
    the last column (and many of the negative ones) should be treated as
    consistent with zero.  } 
\label{tab:constraints}
  \begin{tabular}{|| c || c | c | c | c ||}
\hline  \hline    
 \textbf{Curvature}      & DES Y1                 &  External                 &  DES Y1 + External       & \dchisq \\ \hline
    $\Omega_k$           & $0.16^{+0.09}_{-0.14}$& $0.0023^{+0.0035}_{-0.0030}$ & $0.0020^{+0.0037}_{-0.0032}$&\multirow{1}{*}{$[-0.9, -0.2, -0.1]$} \\\hline\hline
 \textbf{Number Rel.\ Species}     & DES Y1       &  External                 &  DES Y1 + External & \dchisq\\ \hline
    $\Neff$              & $<5.38$                & $<3.32$                   & $<3.28$&\multirow{1}{*}{$[0.2, 0.4, -0.7]$}\\\hline\hline
 \textbf{Dynamical dark energy}    & DES Y1       &  External                 &  DES Y1 + External& \dchisq\\ \hline
    $w_0$                & $ -0.69^{+0.30}_{-0.29}$ & $-0.96^{+0.10}_{-0.08}$     & $-0.95^{+0.09}_{-0.08}$&\multirow{3}{*}{$[-1.9, -0.0, -0.1]$}\\
    $w_a$                & $-0.57^{+0.93}_{-1.11}$  & $-0.31^{+ 0.38}_{-0.52}$    & $ -0.28^{+0.37}_{-0.48}$&\\
    $w_p$                & $-0.91^{+0.19}_{-0.23}$  & $-1.02^{+0.04}_{-0.04} $    & $-1.01^{+0.04}_{-0.04}$&\\\hline\hline
 \textbf{Modified Gravity}         & DES Y1       &  External                 &  DES Y1 + External& \dchisq\\ \hline
    $\Sigma_0$           & $0.43^{+0.28}_{-0.29}$   & $0.26^{+0.14}_{-0.13}$      & $0.06^{+0.08}_{-0.07}$ &\multirow{2}{*}{$[-0.2, -3.4, -0.4]$}\\
    $\mu_0$              & \mbox{---}             & $0.16^{+0.43}_{-0.47}$      & $-0.11^{+0.42}_{-0.46}$&\\\hline
\hline
\end{tabular}
\end{table*}

\begin{figure*}[t] 
\centering
\includegraphics[width=0.35\textwidth]{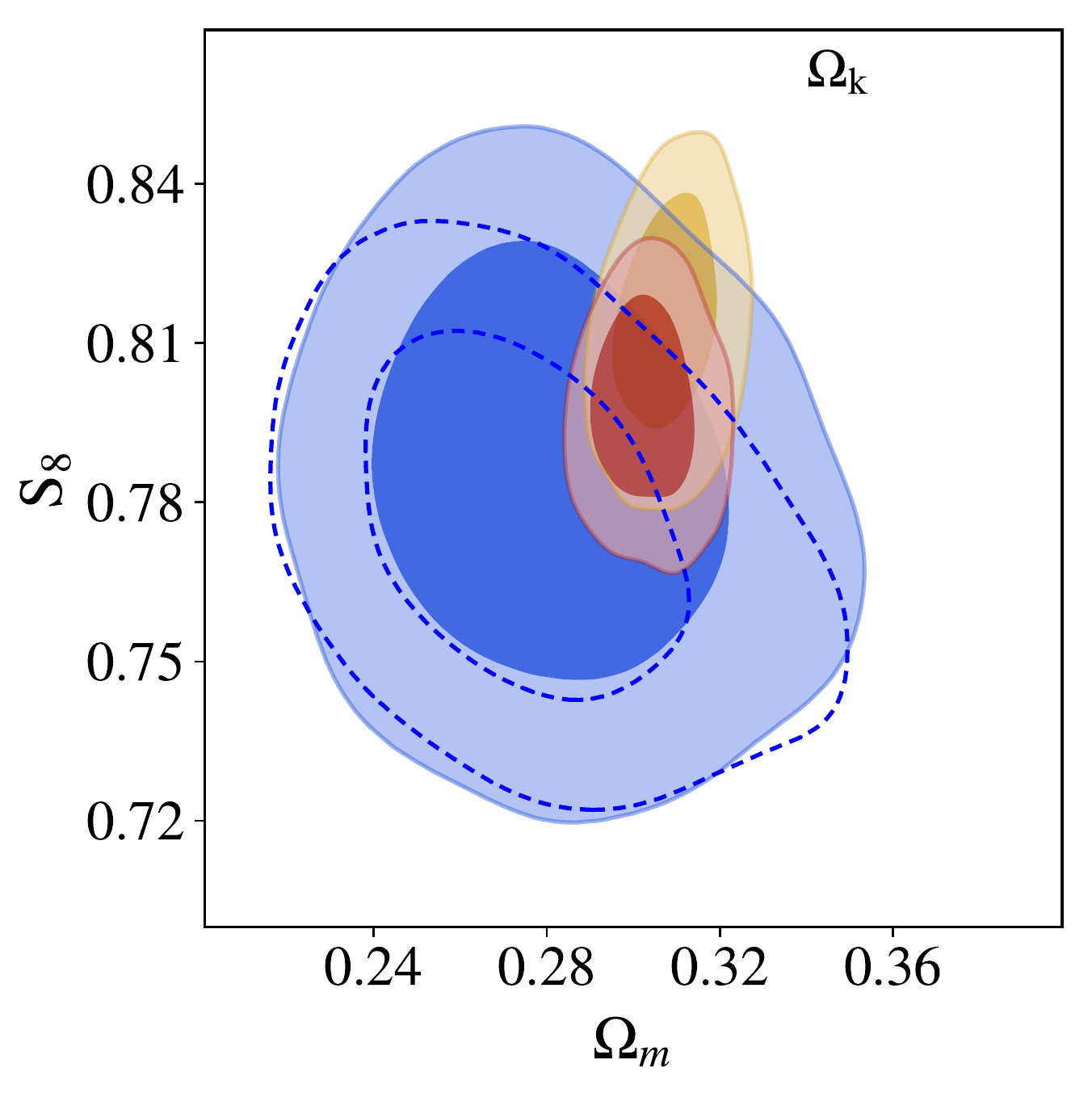}
\includegraphics[width=0.35\textwidth]{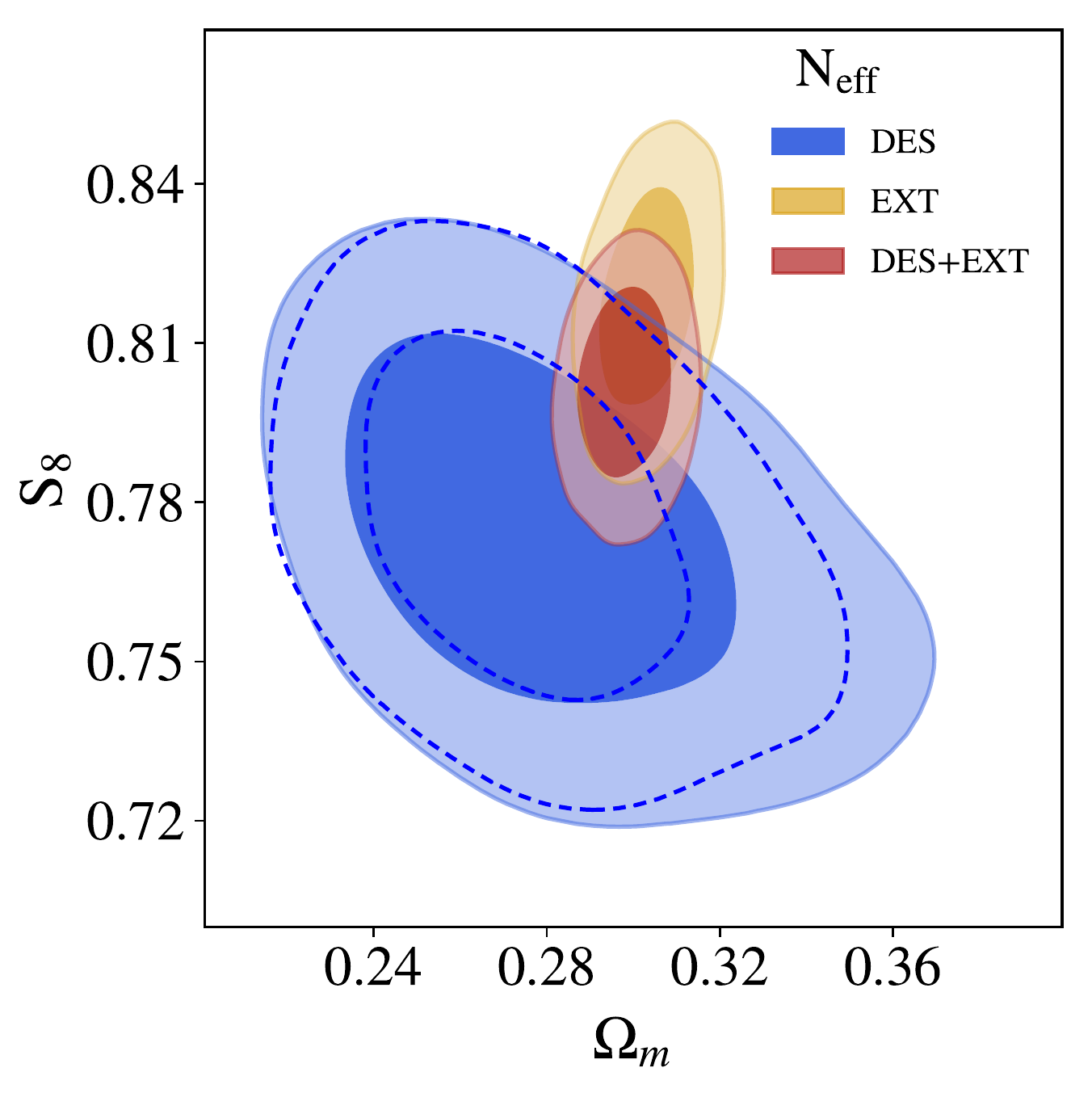}\\
\includegraphics[width=0.35\textwidth]{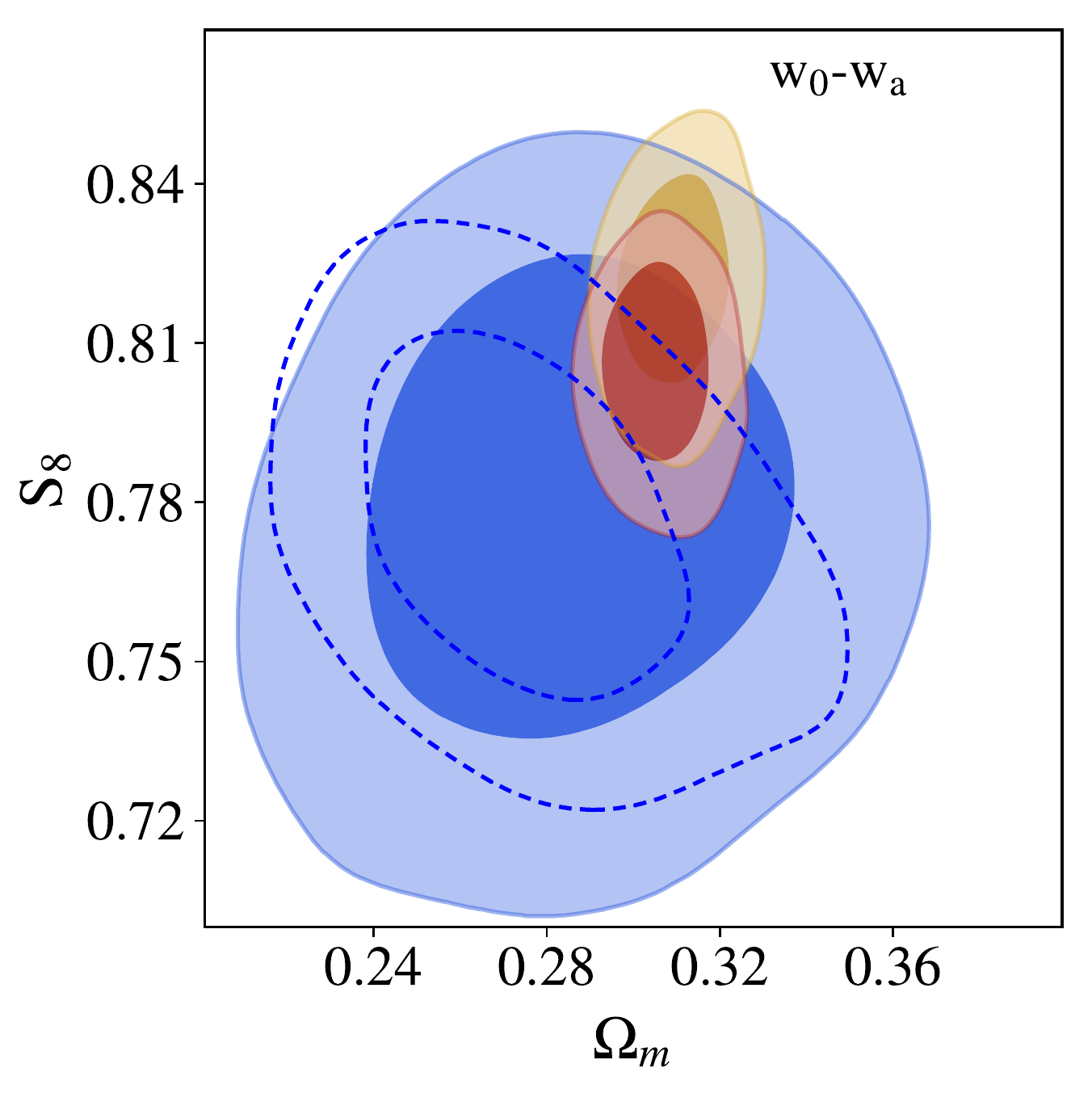}
\includegraphics[width=0.35\textwidth]{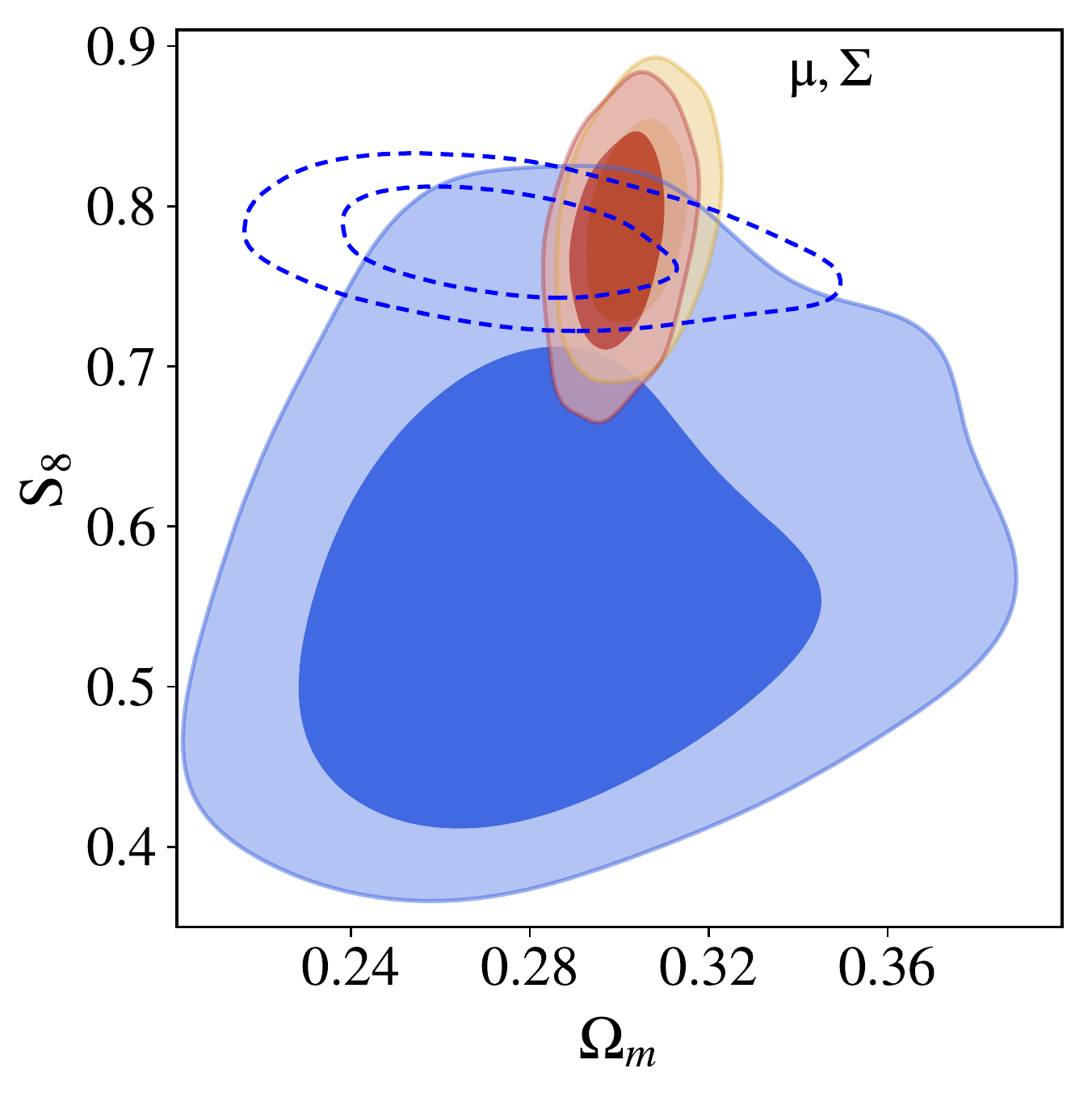}
\caption{Comparison of constraints on the matter density $\Omega_m$ and $S_8$
  to the $\Lambda$CDM case. The panels illustrate how the $\Omega_m$-$S_8$
  constraints broaden and shift as we allow to vary: curvature (\textit{top
    left}), number of relativistic species (\textit{top right}),
  equation-of-state parameters $w_0$ and $w_a$ (\textit{bottom left}), and
  modified gravity parameters $\Sigma_0$ and $\mu_0$ (\textit{bottom
    right}). In each case, the shaded contours denote DES (blue), external
  (yellow), and DES+external (red) constraints. For comparison, in the
  DES-only case we also show the constraints in the $\Lambda$CDM model with
  dashed contours, which are the same in each panel.}
\label{fig:Om_S8}
\end{figure*}

Besides the tighter constraint, DES also pushes $\Sigma_0$ closer to its
$\Lambda$CDM value of zero.  An interesting manifestation of the
multidimensionality of the parameter space is that the DES+external value is
lower than either DES or external alone. This arises because DES favors a
lower amplitude of mass fluctuations than that favored by the external data,
due to the lower amplitude of the lensing signal observed by the DES. Because
the lensing amplitude is proportional to the product $\Sigma_0 S_8$, these two
parameters are highly anticorrelated in DES, and the lensing amplitude
suppression can be accommodated by decreasing either of them.  Since external
data constrain mostly $S_8$ and constrain it to be high, the DES lensing
amplitude is accommodated by shifting $\Sigma_0$ down.


The constraints on the extensions parameters are summarized in Table
\ref{tab:constraints}. The last column in the Table shows the improvement in the goodness-of-fit
between the corresponding best-fit extension and the best-fit $\Lambda$CDM
model, and indicates that none of the extensions are strongly preferred
relative to $\Lambda$CDM.

In Fig.~\ref{fig:Om_S8}, we show the constraints in the
$\Omega_m$-$S_8$ plane for the extended models (solid contours); for
comparison, we also show the $\Lambda$CDM model constraints for DES data alone
(dashed contours which are the same in all panels). The top right corner of
each panel shows which extension the plot is referring to. For $\Omega_k$,
$\Neff$ and $w_0$-$w_a$ extensions, we see that the $\Omega_m$-$S_8$ contour
from DES alone is only modestly increased by marginalization over the
additional nuisance parameter(s). The exception is the modified-gravity case,
where the $\Omega_m$-$S_8$ contour from DES alone is significantly larger and
also pushed to smaller values of $S_8$ because of the amplitude degeneracy
between $\Sigma_0$ and $S_8$.

\Swide{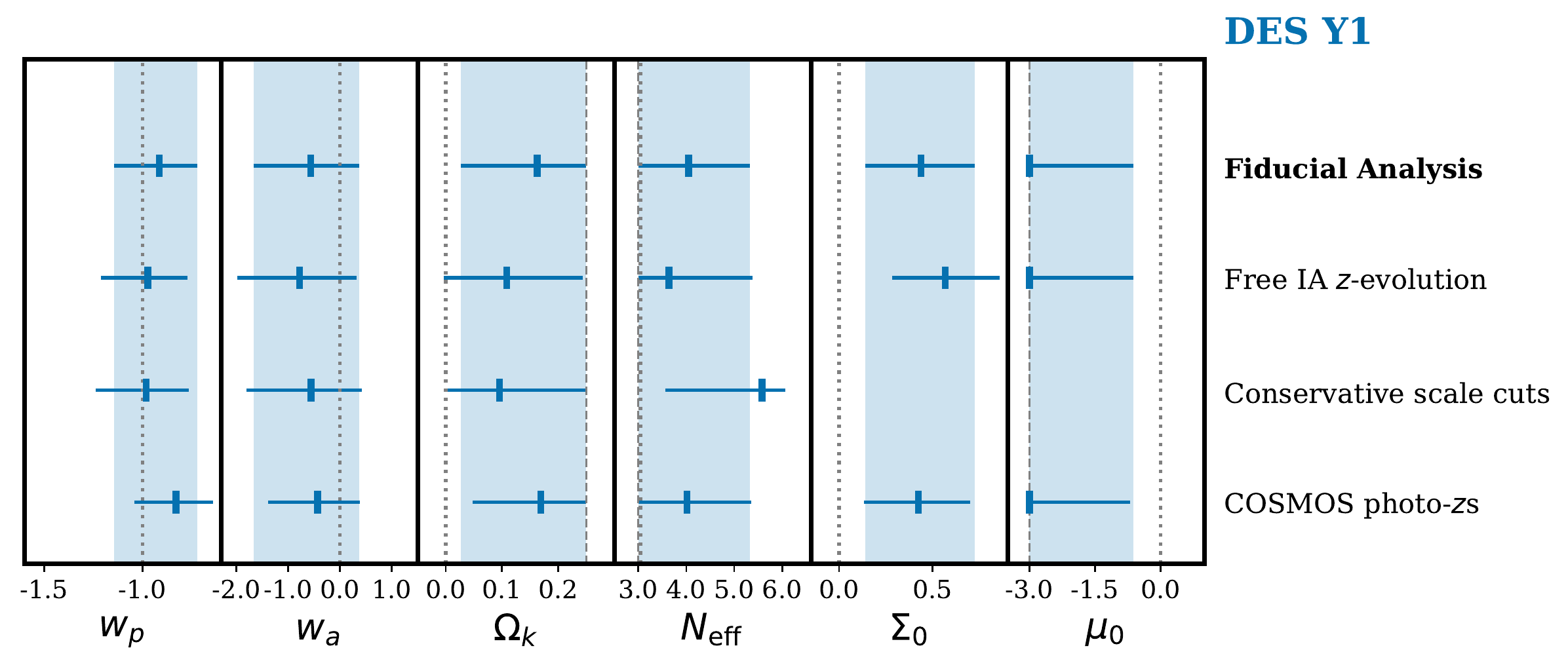}{Impact of changes in modeling
  assumptions to the inferred cosmology, using actual (and not synthesized  as in
  Fig.~\ref{fig:tableplot_extensions_systests}) DES data.  Each column shows
  one of the cosmological parameters describing $\Lambda$CDM extensions.  The
  horizontal error bars show the constraints for each individual change in the
  analysis, listed to the right  of the Figure. The vertical
  shaded band coincides with the horizontal error bars in the
  fiducial-analysis case. The modified-gravity analysis assumes conservative
  scale cuts as a default, so the corresponding test is left blank in the
  table.}

Furthermore, Fig.~\ref{fig:tableplot_extensions_pipelineval} shows the
results of the systematic tests on the analysis assumptions outlined in
Sec.~\ref{sec:valid_analysis}. The top row shows our fiducial constraints on
the extensions parameters presented earlier in this Section, relative to the
corresponding marginalized best-fit value in the same fiducial analysis. The next three
rows show these constraints (still relative to the corresponding best-fit value in
the fiducial analysis): assuming alternative treatment of intrinsic
alignments; the use of conservative scales (except in the modified-gravity
extension which assumes them by default); and adopting alternative photometric
redshifts. The results show no significant biases in the results on the
extensions parameters, providing further support that our modeling is robust
with respect to our modeling of intrinsic alignments, angular scales used, and photometric
redshifts.

We now compare our extended-model cosmological constraints to those obtained
using KiDS-450 \cite{Joudaki:2016kym} shear measurements, and to the Planck
2018 (P18) CMB measurements \cite{Aghanim:2018eyx}. KiDS analysis is similar
to ours in that they use their own shear measurements combined with external
data; one difference is that we use the full $3\times2$ data vector which, in
addition to shear, also includes galaxy clustering and galaxy-galaxy
lensing\footnote{In their extended work \cite{Joudaki:2017zdt}, KiDS
    combine their own shear measurements with galaxy clustering and RSD
    information from 2-degree Field Lensing Survey (2dFLenS) and the Baryon
    Oscillation Spectroscopic Survey (BOSS), effectively using a
    ($3\times2$)-type data vector. Here we choose to compare our DES-only
    results to KiDS-only results presented in Ref.~\cite{Joudaki:2016kym}.  We
    thank Shahab Joudaki for pointing this out.}.  Planck, on the other hand,
uses the DES Y1 shear measurements as an external weak lensing data set,
combining it with their CMB information.  It is important to note that both
KiDS and Planck fix the neutrino mass to $\sum m_\nu=0.06$eV in their baseline
$\Lambda$CDM model, while we vary the neutrino mass as part of the fiducial
model.  Therefore, our cosmological constraints are expected to be weaker, but
more robust with respect to the neutrino mass, than they would be with the
same assumptions as KiDS and P18.

Comparison with KiDS-450 will be necessarily qualitative, given that they do not
quote the numerical values of their constraints on the cosmological parameters.
KiDS do not consider $\Neff$ as one of their extensions, but they do study
curvature, finding some preference for a negative $\Omega_k$ (see their
Fig.~8b), which is in the opposite direction of our mild preference for
positive $\Omega_k$. Their $w_0$-$w_a$ constraint, like ours, is broadly
consistent with the $\Lambda$CDM scenario with values of $-1$ and zero,
respectively. Their phenomenological tests of gravity assumed the ($Q$,
$\Sigma$) parametrization, where $Q_{\rm KiDS} = 1+ 2\Sigma_{\rm
  DES}-\mu_{\rm DES}$ and $\Sigma_{\rm KiDS}=1+\Sigma_{\rm DES}$, so that
general relativity corresponds to their $(Q, \Sigma) = (1, 1)$. They described
each of their functions $Q$ and $\Sigma$ by piecewise constant values across
two bins in scale and two in redshift, so that their analysis included eight
modified-gravity parameters as opposed to two in the present paper. Comparing
DES and KiDS modified-gravity results is therefore not straightforward but we
can study the main trends.  The parameters ($Q_2$, $\Sigma_2$) corresponding
to the modified gravity parameters in the low redshift bin and small-scale
(high-$k$) bin are the best constrained by KiDS and are shown in figure 13 of
\cite{Joudaki:2016kym}.  Much like we see in our own results, KiDS
measurements help constrain $\Sigma$ as it is directly linked to the lensing
potential.  Interestingly, KiDS results are consistent with very positive
values of $Q_2$ (although they are also consistent with the standard value
$Q_2=1$), which corresponds to DES's preference for a positive $\Sigma_0$ and
negative $\mu_0$ shown in the right panel of Fig.~\ref{fig:w0wa_MG}. On the
whole, the different temporal and spatial parametrizations of modified gravity
functions in KiDS and DES Y1, along with other differences in the two
analyses, make detailed comparisons impossible, but the two surveys'
constraints on modified gravity seem in broad agreement.


For comparison with Planck we only consider the modified-gravity case, as this
is the $\Lambda$CDM extension where DES Y1 information appreciably improves
the constraints obtained from Planck and other external data.  P18 constraints
on modified gravity \cite{Aghanim:2018eyx} employ the base parameters $\mu$
and $\eta$, with $\mu_{\rm P18} = 1+\mu_{\rm DES}$ being defined to have the
redshift variation same as ours in Eq.~(\ref{eq:musigform}); they also quote
constraints on $\Sigma_{\rm P18}=1+\Sigma_{\rm DES}$, whose redshift
dependence however does not coincide with ours. Planck considers a similar set
of other data as we do: their SN and RSD datasets are identical to ours; they
use a more extensive selection of BAO data, but their DES information includes
only the weak lensing (shear) information and not the full $3\times2$ data vector as
in the present paper.  Therefore, a somewhat direct although not exact
comparison of the combined constraints between DES Y1 and P18 is possible.  We
refer to Table 7 of \cite{Aghanim:2018eyx} where P18 report constraints from
the combination of Planck and external data, the latter of which includes DES
Y1 shear. The central values of $\Sigma_0$ and $\mu_0$ in our DES+external
analysis are very close to the corresponding values in P18.  Our DES+external
errors on $\Sigma_0$ ($\mu_0$) are about 30\% (80\%) weaker that those in P18,
which is probably chiefly due to our marginalization over neutrino mass, and
possibly also to the aforementioned differences in the selected data sets. On
the whole, the DES and P18 constraints that combine all data are consistent
both mutually and with predictions of general relativity.

The new information that the DES Y1 data contribute to the overall
constraints on modified gravity that we presented in this paper illustrates
that near-future DES data should provide sharp tests of the modified-gravity
paradigm.


\section{Conclusions}
\label{sec:conclu}

The results in this paper extend the work done in the Y1KP \cite{Y1KP} by
analyzing the models beyond flat $\Lambda$CDM and $w$CDM. In Y1KP, we found good
agreement with the standard cosmological-constant dominated universe, and
produced constraints on the matter density and amplitude of mass fluctuations
comparable to those from the Planck satellite. We now extend that work into
four new directions, allowing for: 1) nonzero curvature $\Omega_k$; 2) number
of relativistic species $\Neff$ different from the standard value of 3.046; 3)
time-varying equation-of-state of dark energy described by the parameters
$w_0$ and $w_a$ (alternatively, the values at the pivot redshift $w_p$ and
$w_a$); and 4) modified gravity described by the parameters $\Sigma_0$,
$\mu_0$ that modify the metric potentials.

For the first three of these four extensions, we find that the DES Y1 data
alone are consistent with values of zero curvature, three relativistic
species, and dark energy parameters corresponding to the cosmological constant
model. We also find that DES Y1 data do not significantly improve the existing
constraints which combine the Planck 2015 temperature and polarization
measurements, BAO measurements from SDSS and BOSS, RSD measurements from BOSS,
and type Ia supernova measurements from the Pantheon compilation. When DES Y1
information is combined with that from the external data, the constraints on
curvature are $\Omega_k=0.0020^{+0.0037}_{-0.0032}$, while that on the
dark-energy equation of state pivot value and its variation are
$w_p=-1.01^{+0.04}_{-0.04}$ and $w_a=-0.28^{+0.37}_{-0.48}$, respectively.
The upper bound on the number of relativistic species is $\Neff<3.28 (3.55)$ at the 68\%
(95\%) confidence level from the combination of DES and external data.

DES Y1 alone provides a stronger constraint on the fourth extension of
$\Lambda$CDM that we consider -- modified gravity -- giving $\Sigma_0 =
0.43^{+0.28}_{-0.29}$.  The apparent DES-alone preference for positive
$\Sigma_0$ is consistent with parameter volume effects discussed in
Sec.~\ref{sec:valid_sim}. When combining DES with external data, the
$\Sigma_0$ constraint is shifted downwards with respect to the external-only
constraint, which can be explained by the fact that DES data prefer a lower
lensing amplitude than that predicted by external data in $\Lambda$CDM. Combining DES Y1 with the external
data gives $\Sigma_0 = 0.06^{+0.08}_{-0.07}$ and $\mu_0 =
-0.11^{+0.42}_{-0.46}$, both of which are fully consistent with the
$\Lambda$CDM values $(\Sigma_0, \mu_0)=(0, 0)$.

We applied a suite of validation and null tests both to our analysis and to
our theory modeling; the results of these tests are shown in
Figs.~\ref{fig:tableplot_extensions_systests} and
\ref{fig:tableplot_extensions_pipelineval}. In nontrivial model spaces
such as modified gravity, we compared the results obtained by two
independently developed parameter inference pipelines, {\tt CosmoLike} and
{\tt CosmoSIS}, and also compared the constraints used obtained using two different
samplers, {\tt emcee} and {\tt multinest}. We modeled any remaining
systematics with 20 nuisance parameters, marginalizing over them to get the
constraints on cosmological parameters. Finally, in all cases we applied the
parameter-level blinding procedure, and did not look at the final cosmological
constraints until after unblinding. 
  
The results in this paper also serve to develop the tools necessary to take
advantage of future constraints on these cosmological models by DES. In
particular, the forthcoming analysis of the DES Y3 data, which will contain
information from three times the area of Y1, should provide very interesting
constraints on extensions of the minimal cosmological model including dark
energy and modified gravity.

\section*{Acknowledgments}

We are grateful to the anonymous referee for many useful questions, comments, and
clarifications.

Funding for the DES Projects has been provided by the U.S. Department of Energy, the U.S. National Science Foundation, the Ministry of Science and Education of Spain, 
the Science and Technology Facilities Council of the United Kingdom, the Higher Education Funding Council for England, the National Center for Supercomputing 
Applications at the University of Illinois at Urbana-Champaign, the Kavli Institute of Cosmological Physics at the University of Chicago, 
the Center for Cosmology and Astro-Particle Physics at the Ohio State University,
the Mitchell Institute for Fundamental Physics and Astronomy at Texas A\&M University, Financiadora de Estudos e Projetos, 
Funda{\c c}{\~a}o Carlos Chagas Filho de Amparo {\`a} Pesquisa do Estado do Rio de Janeiro, Conselho Nacional de Desenvolvimento Cient{\'i}fico e Tecnol{\'o}gico and 
the Minist{\'e}rio da Ci{\^e}ncia, Tecnologia e Inova{\c c}{\~a}o, the Deutsche Forschungsgemeinschaft and the Collaborating Institutions in the Dark Energy Survey. 

The Collaborating Institutions are Argonne National Laboratory, the University of California at Santa Cruz, the University of Cambridge, Centro de Investigaciones Energ{\'e}ticas, 
Medioambientales y Tecnol{\'o}gicas-Madrid, the University of Chicago, University College London, the DES-Brazil Consortium, the University of Edinburgh, 
the Eidgen{\"o}ssische Technische Hochschule (ETH) Z{\"u}rich, 
Fermi National Accelerator Laboratory, the University of Illinois at Urbana-Champaign, the Institut de Ci{\`e}ncies de l'Espai (IEEC/CSIC), 
the Institut de F{\'i}sica d'Altes Energies, Lawrence Berkeley National Laboratory, the Ludwig-Maximilians Universit{\"a}t M{\"u}nchen and the associated Excellence Cluster Universe, 
the University of Michigan, the National Optical Astronomy Observatory, the University of Nottingham, The Ohio State University, the University of Pennsylvania, the University of Portsmouth, 
SLAC National Accelerator Laboratory, Stanford University, the University of Sussex, Texas A\&M University, and the OzDES Membership Consortium.

Based in part on observations at Cerro Tololo Inter-American Observatory, National Optical Astronomy Observatory, which is operated by the Association of 
Universities for Research in Astronomy (AURA) under a cooperative agreement with the National Science Foundation.

The DES data management system is supported by the National Science Foundation under Grant Numbers AST-1138766 and AST-1536171.
The DES participants from Spanish institutions are partially supported by MINECO under grants AYA2015-71825, ESP2015-88861, FPA2015-68048, SEV-2012-0234, SEV-2016-0597, and MDM-2015-0509, 
some of which include ERDF funds from the European Union. IFAE is partially funded by the CERCA program of the Generalitat de Catalunya.
Research leading to these results has received funding from the European Research
Council under the European Union's Seventh Framework Program (FP7/2007-2013) including ERC grant agreements 240672, 291329, and 306478.
We  acknowledge support from the Australian Research Council Centre of Excellence for All-sky Astrophysics (CAASTRO), through project number CE110001020.

This manuscript has been authored by Fermi Research Alliance, LLC under Contract No. DE-AC02-07CH11359 with the 
U.S. Department of Energy, Office of Science, Office of High Energy Physics. 
The United States Government retains and the publisher, by accepting the article for publication, 
acknowledges that the United States Government retains a non-exclusive, paid-up, irrevocable, 
world-wide license to publish or reproduce the published form of this manuscript, or allow 
others to do so, for United States Government purposes.

This research used resources of the National Energy Research Scientific
Computing Center, a DOE Office of Science User Facility supported by the
Office of Science of the U.S. Department of Energy under Contract
No. DE-AC02-05CH11231.  Some calculations in this work were performed on the
CCAPP condo of the Ruby Cluster and the Owens Cluster at the Ohio
Supercomputer Center \cite{OhioSupercomputerCenter1987}. This work also used
the Bridges system, which is supported by NSF Award No.\ ACI-1445606, at the
Pittsburgh Supercomputing Center (PSC)
\citep{Nystrom:2015:BUF:2792745.2792775}.

\bibliography{refs}
 
\appendix

\section{{\tt CosmoSIS} and   {\tt CosmoLike}  comparison in the context of testing gravity}
 \label{app:mg_cosmosislike}

In the course of our analyses, we have compared the parameter estimation code
{\tt CosmoSIS} \cite{Zuntz:2014csq} used in Y1KP to the {\tt CosmoLike}
\cite{Krause:2016jvl} code.
The two codes show excellent agreement within the statistical error bars as
shown in \cite{methodpaper}, giving us confidence that our analysis pipeline
is robust.  In the present paper, we have made substantial modifications (as
described below) to the {\tt CosmoSIS} pipeline, which we use as our principal
analysis tool, for the case of the parametrized test of gravity.  In order to
validate the {\tt CosmoSIS} pipeline, we compare its results to those from {\tt
  CosmoLike}.  We first give a brief description of the {\tt CosmoSIS} and
{\tt CosmoLike} pipelines as applied to the case of parametrized tests of
gravity and then show the results of this comparison.


The {\tt CosmoSIS} pipeline has been used in Y1KP and is further described in
\cite{methodpaper}.  To apply {\tt CosmoSIS} to modified-gravity model analysis,
we adopted the publicly available code {\tt MGCAMB}, instead of CAMB, for the computation of the
matter and CMB power spectra. {\tt MGCAMB} doesn't come with the
parametrization of modified gravity identical to ours, so we analytically
translate our ($\Sigma_0$, $\mu_0$) parameters into {\tt MGCAMB}'s ($\gamma$, $\mu$).
We use the January 2012 version of {\tt MGCAMB} to perform the systematics
checks, and the more recent 2015 version for the constraints on real data.
We further modify the part of the pipeline that projects the matter power
spectrum into clustering and weak lensing power spectra in order to account for the
modified-gravity parameters.

While {\tt MGCAMB} embedded in {\tt CosmoSIS} pipeline modifies the perturbed
gravitational potentials and the CMB source functions, {\tt CosmoLike}
directly modifies the lensing kernel with $\Sigma_0$ and the growth factor
with $\mu_0$.  The two pipelines should be equivalent except for the ISW
effect which is implemented in {\tt MGCAMB} and not in {\tt CosmoLike}.  We
therefore expect significant differences in the low multipole part of the CMB
power spectra, but not elsewhere.


First, we have checked that the weak lensing and clustering observables $\xi_{\pm}(\theta)$, $\gamma_t(\theta)$, $w(\theta)$ as computed by {\tt CosmoSIS} and {\tt CosmoLike} agree well
(difference well below the DES Y1 error bars) for a few sets of $(\Sigma_0, \mu_0)$ values.

\begin{figure}[] 
\centering
\includegraphics[width=0.5\textwidth]{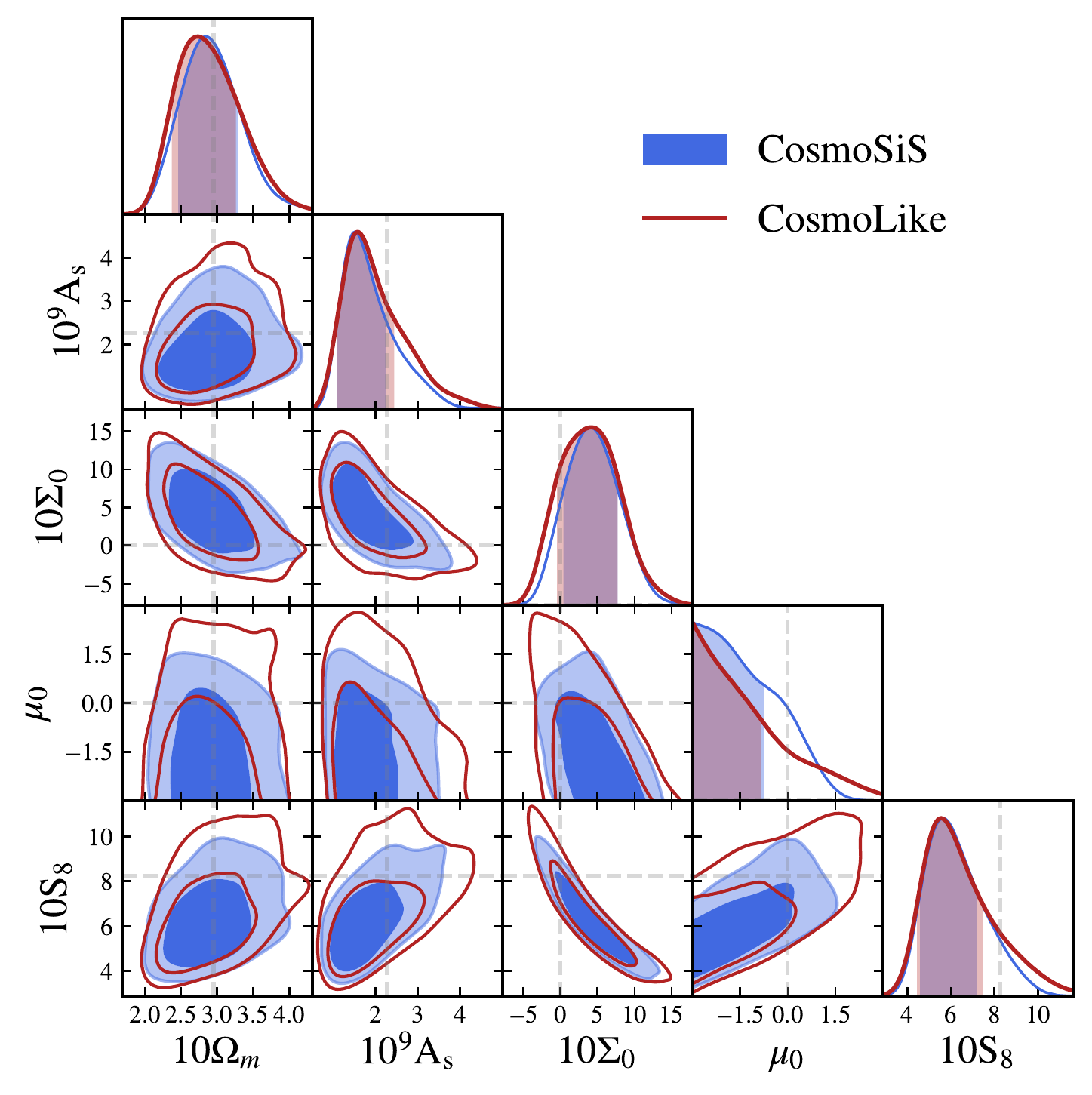} 
\caption{Constraints on $\Omega_m$, $A_s$, $S_8$, $\Sigma_0$ and $\mu_0$ using DES
  Y1 synthetic data for {\tt CosmoSIS} (blue contours) and {\tt CosmoLike} 
  (red).  }
\label{fig:cosmosislike}
\end{figure}

Second, we explicitly test the consistency of the {\tt CosmoSIS} and {\tt
  CosmoLike} pipelines, comparing the constraints they report in the full
parameter space.  To do this we use the {\tt emcee} sampler on synthetic  DES Y1 data,
varying the parameters over the prior ranges used in the main
analysis. Fig. \ref{fig:cosmosislike} shows the results for {\tt CosmoSIS}
(blue) and {\tt CosmoLike} (red) for a subset of the parameters, namely
$\Omega_m$, $A_s$, $\sigma_8$, $\Sigma_0$ and $\mu_0$.  The two pipelines give
similar results, with the 1$\sigma$ contours agreeing very well for all
parameters plotted.  However the 2$\sigma$ contours are wider for {\tt
  CosmoLike} in some cases, specifically for pairs of parameters including the
modified gravity parameter $\mu_0$.  This difference is most striking in the
($\Sigma_0$,$\mu_0$) plane.  This is due to {\tt MGCAMB} failing for sets of
($\Sigma_0$,$\mu_0$) in extreme areas.  Thanks to its implementation of the
modified gravity parameters, {\tt CosmoLike} does not have this issue and is
therefore able to explore a wider range of ($\Sigma_0$,$\mu_0$).  In
particular for this case of synthetic DES Y1 data, the 2-$\sigma$ contours as
derived from {\tt CosmoLike} extends to more positive $\mu_0$ than in {\tt
  CosmoSIS}. This partially explains the constraints on $\mu_0$ from the real
DES Y1 data shown in Fig. \ref{fig:w0wa_MG} in the area where $\mu_0$ is very
positive.  However we note that using the more constraining data sets place us
far from these more extreme areas and therefore these results are safe from
this issue.
 
\end{document}